\documentclass[aps,prd,superscriptaddress,showkeys,showpacs,twocolumn,a4paper]{revtex4-1}
\usepackage{ulem}
\usepackage{color}
\usepackage{graphicx}
\usepackage{hyperref}   
\usepackage{appendix}
\usepackage{epsfig}
\usepackage{epstopdf}
\usepackage{amsmath}
\usepackage{subfig}
\usepackage{comment}
\usepackage[dvipsnames]{xcolor}
\newcommand{\be}{\begin{equation}}
\newcommand{\ee}{\end{equation}}
\newcommand{\ba}{\begin{eqnarray}}
\newcommand{\ea}{\end{eqnarray}}

\begin{document}

\title{First order dissipative hydrodynamics and viscous corrections to the entropy four-current from an effective covariant kinetic theory}

\author{Samapan Bhadury}
\email{samapan.bhadury@niser.ac.in}
\affiliation{School of Physical Sciences, National Institute of Science Education and Research, HBNI, Jatni-752050, India}
\author{Manu Kurian}
\email{manu.kurian@iitgn.ac.in}
\affiliation{Indian Institute of Technology Gandhinagar, Gandhinagar-382355, Gujarat, India}
\affiliation{Department of Physics, McGill University, 3600 University Street, Montreal, QC, H3A 2T8, Canada}
\author{Vinod Chandra}
\email{vchandra@iitgn.ac.in}
\affiliation{Indian Institute of Technology Gandhinagar, Gandhinagar-382355, Gujarat, India}
\author{Amaresh Jaiswal}
\email{a.jaiswal@niser.ac.in}
\affiliation{School of Physical Sciences, National Institute of Science Education and Research, HBNI, Jatni-752050, India}

%%%%%%%%%%%%%%%%%%%%%%%%%%%%%%%%%%%%%%%%%%%%%%%%%%%%%%%%%%%%%%%%%%%%%

\begin{abstract}
The first order hydrodynamic evolution equations for the shear stress tensor, the bulk viscous pressure and the charge current have been studied for a system of quarks and gluons, with a non-vanishing quark chemical potential and finite quark mass. The first order transport coefficients have been obtained by solving an effective Boltzmann equation for the grand-canonical ensemble of quasiquarks and quasigluons. We adopted temperature dependent effective fugacity for the quasiparticles to encode the hot QCD medium effects. The non-trivial energy dispersion of the quasiparticles induces mean field contributions to the transport coefficients whose origin could be directly related to the realization of  conservation laws from  the effective kinetic theory. Both the QCD equation of state and chemical potential are seen to have a significant impact on the quark-gluon plasma evolution. The shear and bulk viscous corrections to the entropy-four current have been investigated in the framework of the effective kinetic theory. The effect of viscous corrections to the entropy density have been quantified in the case of one dimensional boost-invariant expansion of the system. Further, the first order viscous corrections to the time evolution of temperature along with the description of pressure anisotropy and Reynolds number of the system have been explored for the longitudinal boost-invariant expansion.
\end{abstract}

%%%%%%%%%%%%%%%%%%%%%%%%%%%%%%%%%%%%%%%%%%%%%%%%%%%%%%%%%%%%%%%%%%%%%

%\pacs{12.38.Mh, 13.40.-f, 05.20.Dd, 25.75.-q}

\maketitle

%%%%%%%%%%%%%%%%%%%%%%%%%%%%%%%%%%%%%%%%%%%%%%%%%%%%%%%%%%%%%%%%%%%%%
 
\section{Introduction}

High energy heavy-ion collision (HIC) experiments in Relativistic Heavy Ion Collider (RHIC) and Large Hadron Collider (LHC) have realized the existence of a new state of matter-the quark-gluon plasma (QGP), which turned out to be more like a near-perfect fluid~\cite{STAR, Aamodt:2010pb, Heinz:2008tv}. Relativistic hydrodynamics has been successfully employed to describe the space-time evolution of the QGP; see Refs.~\cite{Jeon:2015dfa, Florkowski:2017olj, Florkowski:2019qdp, Heinz:2013th, Braun-Munzinger:2015hba, Jaiswal:2016hex} for recent reviews. On the other hand, the input parameters, such as the equation of state and transport coefficients, have been estimated from the microscopic theories. The inclusion of dissipative effects in the QGP evolution is significant for explaining the quantitative behavior of experimental observables in the HIC, i.e., collective flow, transverse momentum spectra, etc. \cite{Denicol:2010xn, Baier:2006um, Baier:2007ix, Denicol:2012cn, Bhalerao:2013pza}. The theoretical explanation of the hadron elliptic flow at RHIC with dissipative hydrodynamic evolution provide the evidence of transport processes in the QCD medium \cite{Luzum:2008cw}. The relevance of the transport process in the HIC is reconfirmed in~\cite{ALICE:2016kpq, Adam:2016izf, Abelev:2013cva, Adam:2016nfo}.

There have been various approaches/attempts for the estimation of the transport parameters of the hot QCD medium~\cite{Arnold:2000dr, Arnold:2003zc, Meyer:2007ic, Meyer:2007dy, Bluhm:2011xu, Deb:2016myz, Ghosh:2015mda, Mitra:2016zdw, Jaiswal:2014isa, Florkowski:2015lra}. To explore the relative significance of transport parameters, their ratios have been studied in recent works~\cite{Marty:2013ita, Mitra:2017sjo}. The quantitative estimation of shear viscosity from experiments has been widely investigated in several works~\cite{Niemi:2011ix, Niemi:2012ry, Romatschke:2007mq, Song:2008hj, Song:2010aq, Song:2011qa, Song:2011hk, Schenke:2010rr, Gale:2012rq}. In parallel, there have been some attempts to study the effect of bulk viscosity in the evolution of the QGP~\cite{Ryu:2015vwa, Huang:2010sa, Denicol:2014vaa, Dobado:2011qu, Bluhm:2010qf}. Notably, the effect of dissipative charge current has received less attention compared to the viscous coefficients in the framework of dissipative hydrodynamics. This can be attributed to the fact that the net baryon number and chemical potential are insignificant in the very high energetic collisions. However, for the lower collision energies probed in the RHIC beam energy scan and for upcoming experiments at Facility for Antiproton and Ion Research (FAIR), the baryon chemical potential can no longer be neglected. In addition to the effects of the chemical potential, the finite quark mass corrections are also significant in the evolution of the QGP in this context. This sets the motivation to investigate the hydrodynamic evolution of the QGP with a non-vanishing baryon chemical potential and finite quark mass. 

The description of the QCD medium evolution requires the knowledge of the microscopic description of thermodynamic quantities of the medium along with the appropriate momentum distribution functions of its effective degrees of freedom (quasi-quarks/antiquarks and quasi-gluons). To that end, encoding the thermal medium effects in the hot QCD equations of state (computed within lattice QCD or Hard Thermal Loop theory) in terms of quasiparton degrees of freedom with nontrivial dispersion relations has turned out to be a viable approach. The quasiparticle description of thermodynamic and transport properties of the hot QCD/QGP medium have been investigated in several works~\cite{Mitra:2016zdw, Mitra:2017sjo, Kurian:2018dbn, Kurian:2017yxj, Kurian:2018qwb, Romatschke:2011qp, Tinti:2016bav, Alqahtani:2016rth, Alqahtani:2017jwl, Rozynek:2018tev}. In the current analysis, we utilize the effective fugacity quasiparticle model (EQPM)~\cite{Chandra:2011en, Chandra:2007ca} for the effective microscopic description of the QGP. The microscopic framework for the estimation of transport coefficients in the current analysis is done within the covariant kinetic theory approach using the relativistic Boltzmann equation. We employ the Chapman-Enskog like iterative method to solve the relativistic transport equation with relaxation time approximation (RTA) for the collision kernel, along with a mean field term arising from the quasi-particle description of the QGP.
The mean field term in the effective covariant kinetic theory with the EQPM can be realized from the conservation laws as described in the Ref.~\cite{Mitra:2018akk}. The goal of the current analysis is to investigate these mean field corrections to the dissipative quantities with non-vanishing baryon chemical potential and quark mass, within the EQPM. The relative behavior of different dissipative processes can be estimated with the respective ratios of their transport coefficients in the light of the mean field contributions and finite quark chemical potential. We study the viscous corrections to the time evolution of temperature, pressure anisotropy, and Reynolds number by analyzing the boost invariant longitudinal expansion. The four-vector entropy current may have contributions from shear viscous tensor and bulk viscous pressure for a viscous fluid~\cite{El:2009vj,Chattopadhyay:2014lya}. We estimate the viscous corrections to the entropy current from  the effective kinetic theory and demonstrate the significance of these dissipative effects in the case of longitudinal expansion. These aspects are crucial in the investigation of the hydrodynamic evolution of the QGP from the covariant effective kinetic theory.

The manuscript is organized as follows. The mathematical formulation of the first order dissipative hydrodynamic evolution equations from the EQPM covariant kinetic theory along with the description of longitudinal Bjorken flow and viscous corrections to entropy four-current are presented in section II. Section III deals with the results and discussions of the present analysis. Finally, in section IV, the conclusion and outlook have been presented.

\textbf{Notations and Conventions:} In this article, we have used the following notations and conventions. The quantity $u_\mu$ is the fluid velocity (normalized to unity) and in the fluid rest frame $u^{\mu}=(1,0,0,0)$. The metric tensor is taken to be $g^{\mu\nu}=\text{diag}(1, -1, -1, -1)$. The subscript $k$ used in the manuscript implies the particle species, $k=(g, q,\bar{q})$, where $g$, $q$ and $\bar{q}$ denotes gluons, quarks and antiquarks, respectively. The quantity $g_k$ represents the degeneracy factor of the $k$-th species. We choose the appropriate gluon and quark/antiquark degeneracy factors respectively as $g_g=N_s\times (N_c^2-1)$ and $g_q=N_s\times N_c\times N_f$, where $N_f=3$ is the number of flavors, $N_s=2$ is the spin degrees of freedom and $N_c=3$ is the number of colors.
%%%%%%%%%%%%%%%%%%%%%%%%%%%%%%%%%%%%%%%%%%%%%%%%%%%%%%%%%%%%%%%%%%%%%

\section{Formalism}

The formalism for the estimation of dissipative hydrodynamic evolution of the QGP consists of the quasiparticle modeling followed by the setting up of the effective covariant kinetic theory of the system away from equilibrium. The current analysis is based on a covariant kinetic theory for hot QCD medium  recently developed by Chandra and Mitra~\cite{Mitra:2018akk} employing the  effective fugacity quasiparticle model~\cite{Chandra:2011en, Chandra:2007ca}. Here, we have extended the approach to investigate the transport properties of the hot QCD medium with finite quark chemical potential and quark-antiquark masses. 
There are several other quasiparticle models present in the literature to describe hot QCD medium~\cite{kamf, Peshier:1995ty, DElia:1997sdk, DElia:2002hkf,Castorina:2007qv, Castorina:2005wi,Bannur:2006js, Koothottil:2018akg, Dumitru:2001xa, Fukushima:2003fw, Ghosh:2006qh,Su:2014rma, Florkowski:2015dmm, Bandyopadhyay:2015wua}.

%%%%%%%%%%%%%%%%%%%%%%%%%%%%%%%%%%%%%%%%%%%%%%%%%%%%%%%%%%%%%%%%%
\subsection{QCD thermodynamics and the effective covariant kinetic theory with finite chemical potential}

Realizing the hot QCD medium as a Grand-canonical ensemble, the EQPM interprets the hot QCD equation of states (EoS) with quasigluon and quasiquark/antiquark effective fugacities. Here, we have considered the $(2+1)-$flavor lattice QCD EoS for the effective description of QGP~\cite{Cheng:2007jq, Borsanyi:2013bia}. The EQPM energy-momentum tensor can be defined in terms of dressed momenta $\vec{\Tilde{p_k}}$ of k-th particle species and takes the following form~\cite{Mitra:2018akk},
\begin{align}\label{1}
T^{\mu\nu}(x)=&~\sum_{k}g_k\int{d\Tilde{P}_k\,\Tilde{p}_k^{\mu}\,\Tilde{p}_k^{\nu}\,f_k(x,\Tilde{p}_{k})}\nonumber\\
&+\sum_{k}\delta\omega_k\,g_k\int{d\Tilde{P}_k\,\frac{\langle\Tilde{p}_k^{\mu}\,\Tilde{p}_k^{\nu}\rangle}{E_k}\, f_k(x,\Tilde{p}_{k})},
\end{align}
where $f_k(x,\Tilde{p}_{k})$ is the quasiparton distribution function, $\delta\omega_k$ is the modified part of the quasiparicle dispersion relation, $\langle\Tilde{p}_k^{\mu}\,\Tilde{p}_k^{\nu}\rangle\equiv\frac{1}{2}(\Delta^{\mu}_{\alpha}\Delta^{\nu}_{\beta}+\Delta^{\mu}_{\beta}\Delta^{\nu}_{\alpha})\,\tilde{p}_k^{\alpha}\,\Tilde{p}_k^{\beta}$ and $d\Tilde{P}_k\equiv\frac{d^3\mid\vec{\Tilde{p}}_k\mid}{(2\pi)^3\omega_{k}}$ is the momentum integral measure with $\omega_k$ as the quasiparticle energy. In the above equation, $\Tilde{p}_k^{\mu}$ is the ``dressed" four-momentum of particles of species $k$ defined later in terms of bare particle four momentum $p^\mu_k$ and effective fugacity $z_k$. We consider nonzero quark mass $m_q$ of different flavor (with $m_u=3$ MeV, $m_d=5$ MeV and $m_s=100$ MeV for up, down and strange quarks, respectively) and energy $E_k={\sqrt{\mid\vec{\Tilde{p}}_k\mid^2+m_q^2}}$ for quarks/antiquarks whereas for gluons $E_k=\mid\vec{\Tilde{p}}_k\mid$. 

The covariant form of EQPM parton distribution functions in equilibrium, with a non-zero baryon chemical potential $\mu_q$ can be defined as 
\begin{align}
f^0_q &=\frac{z_q \exp{[-\beta (u\!\cdot\! p_q - \mu_q)]}}{1 + z_q\exp{[-\beta (u\!\cdot\! p_q - \mu_q)]}}, \label{2}\\
f^0_{\bar{q}} &=\frac{z_{\bar{q}} \exp{[-\beta (u\!\cdot\! p_{\bar{q}} + \mu_q)]}}{1 + z_{\bar{q}}\exp{[-\beta (u\!\cdot\! p_{\bar{q}} + \mu_q)]}}, \label{3}\\
f^0_g &=\frac{z_g \exp{[-\beta\, u\!\cdot\! p_g]}}{1 - z_g\exp{[-\beta\, u\!\cdot\! p_g]}}, \label{4}
\end{align}
where we define the scalar product $u\!\cdot\! p\equiv u_\mu\,p^\mu$ and the inverse temperature $\beta\equiv1/T$. 
The dispersion relation relates the quasiparticle (dressed) four-momenta $\Tilde{p}_k^{\mu}$ and the bare particle four-momenta $p_k^{\mu}$ as follows,
\begin{equation}\label{5}
\Tilde{p_k}^{\mu} = p_k^{\mu}+\delta\omega_k\, u^{\mu}, \qquad
\delta\omega_k= T^{2}\,\partial_{T} \ln(z_{k}).
\end{equation}
The zeroth component of the four-momenta is modified as,
\begin{equation}\label{6}
\Tilde{p_k}^{0}\equiv\omega_{k}=E_k+\delta\omega_k.
\end{equation}
The quantities $z_q$, $z_{\bar{q}}$ and $z_g$ denote the temperature dependent effective fugacity parameter of the quarks, anti-quarks and gluons, respectively, that encode the hot QCD medium effects in the quasiparticle description of the QGP. The effective fugacities are not related with any conserved number current in the hot QCD medium and the fugacity parameter for quark and antiquark is same, $i.e.$, $z_q=z_{\bar{q}}$ in the EQPM description of the QGP~\cite{Chandra:2011en, Chandra:2007ca}. Therefore, in the rest of this article, we denote the fugacity parameter for both quasiquark and antiquark by $z_q$. 
Since the effective fugacities are not related with any conserved number current in the QGP medium, the temperature dependence of the effective fugacity parameter remain unaltered with the finite chemical potential, as discussed in the Ref.~\cite{Mitra:2017sjo}. 
As expected, in the limit of vanishing quark chemical potential, i.e., $\mu_q=0$, the equilibrium distribution functions for quasiquark and anti-quark becomes identical $f^0_q\equiv f^0_{\bar{q}}$.

Next, we focus on the net baryon four-current $N^{\mu}$ which is defined as the difference of baryon and anti-baryon four-current~\cite{vogt, Jaiswal:2015mxa}. The quasiparticle description of the flow in terms of dressed momenta has the following form~\cite{Mitra:2018akk}
\begin{align}\label{7}
N^{\mu}(x)=&~ g_q\!\int\! d\Tilde{P}_q\,\Tilde{p}_q^{\mu}\left[f_q(x,\Tilde{p}_{q}) - f_{\bar{q}}(x,\Tilde{p}_{q})\right] \nonumber\\
&+\delta\omega_q\,g_q \!\int\! d\Tilde{P}_q\,\frac{\langle\Tilde{p}_q^{\mu}\rangle}{E_q}\! \left[f_q(x,\Tilde{p}_{q}) - f_{\bar{q}}(x,\Tilde{p}_{q})\right],
\end{align}
where $\langle\Tilde{p}^{\mu}_q\rangle\equiv\Delta^\mu_\nu\,\Tilde{p}^{\nu}_q$ is the irreducible tensor of rank one. Note that the quark and antiquark four-momentum is  same $i.e.$  $\Tilde{p}_q^{\mu}=\Tilde{p}_{\bar{q}}^{\mu}$ since both sectors have identical mass $m_q$.  Note that the finite density correction to the effective fugacity leads to a higher order correction in $\mu_q$ in the description of $N^{\mu}$. The relevant thermodynamic quantities such as energy density, pressure, the speed of sound and number density at finite $\mu$ can be obtained from their basic thermodynamic definitions~\cite{Haque:2014rua}. 

From Eq.~(\ref{1}), we obtain the expression of the energy density $\varepsilon$ and the pressure $P$, respectively, within the EQPM by using the following definitions
\begin{align}\label{8}
\varepsilon \equiv u_{\mu}u_{\nu}T^{\mu\nu}, \qquad 
P \equiv -\frac{1}{3}\Delta_{\mu\nu}T^{\mu\nu},
\end{align}
along with the matching condition $\varepsilon=\varepsilon_0$ and $n=n_0$ where the subscript `$0$' represents equilibrium quantities. In the case of finite quark mass, $m_q$, and non-vanishing baryon chemical potential, $\mu_q$, the total energy density and pressure can be expressed in terms of modified Bessel function of second kind, $K_n(y)$, and PolyLog functions as
\begin{align}\label{9}
\varepsilon =&~\sum_{l=1}^\infty g_q\frac{y^4 T^4 (-1)^{l-1} z_q^l\cosh(\alpha l)}{8\pi^2}\bigg[K_4(ly)-K_0(ly)\nonumber\\
&+\frac{2~\delta\omega_q}{yT}\Big(K_3(ly)-K_1(ly)\Big)\bigg] + g_g\frac{T^3}{\pi^2} \nonumber\\
&\times \bigg[ 3T\,\mathrm{PolyLog}~[4,z_g] + \delta\omega_g\mathrm{PolyLog}~[3,z_g] \bigg],
\end{align}
and
\begin{align}\label{10}
P =&~\sum_{l=1}^\infty g_q\frac{y^2 T^4(-1)^{l-1}z_q^l~ \cosh(\alpha l)}{\pi^2\,l^2}K_2(ly) \nonumber\\
& + g_g \frac{T^4}{\pi^2} \mathrm{PolyLog}~[4,z_g],
\end{align}
where $y\equiv\beta\, m_q$ and $\alpha\equiv\beta\,\mu_q$. In the $z_q\to1$, i.e., $\delta\omega_q\to0$ limit, the expressions for energy density and pressure  given in Eqs.~(\ref{9})~and~(\ref{10}), respectively, reduce to that obtained in Ref.~\cite{Florkowski:2015lra}. Similarly, one can obtain the quasiparticle net baryon density using the definition $n\equiv u_\mu N^{\mu}$ along with the matching conditions. The net baryon density is then given by
\begin{equation}\label{11}
n = \sum_{l=1}^\infty g_q\frac{2~y^2 T^3(-1)^{l-1} z_q^l\,\sinh(\alpha l)}{\pi^2\,l}K_2(ly).
\end{equation}
Note that in the limit of vanishing chemical potential, $\alpha\to0$, net baryon density disappears.

From Eqs.~(\ref{9})~and~(\ref{10}), one can obtain the results for massless case by setting $m_q=0$. In this case, we obtain the EQPM energy density and pressure of the hot QGP with a non-zero $\mu_q$ in terms of PolyLog functions and have the following form,
\begin{align}\label{12}
\varepsilon =&~\frac{3T^4}{\pi^2}\bigg[g_g\mathrm{PolyLog}~[4,z_g]-g_q\Big( \mathrm{PolyLog}~[4,-z_qe^{-\alpha}] \nonumber\\
&+\mathrm{PolyLog}~[4,-z_qe^{\alpha}]\Big)\bigg] + \frac{\delta\omega_g~g_gT^3}{\pi^2} \mathrm{PolyLog}~[3,z_g] \nonumber\\
&-\frac{\delta\omega_q~g_qT^3}{\pi^2}\Big(\mathrm{PolyLog}~[3,-z_q e^{\alpha}] \nonumber\\
&+\mathrm{PolyLog}~[3,-z_qe^{-\alpha}]\Big),
\end{align} 
and
\begin{align}\label{13}
P =& ~ \frac{T^4}{\pi^2}\bigg[g_g\mathrm{PolyLog}~[4,z_g] - g_q\Big(\mathrm{PolyLog}~[4,-z_qe^{\alpha}] \nonumber\\
&+\mathrm{PolyLog}~[4,-z_qe^{-\alpha}]\Big)\bigg].
\end{align}
Similarly, the net baryon density in the massless limit takes the following form,
\begin{align}\label{14}
n = \frac{g_qT^3}{\pi^2}\bigg(\mathrm{PolyLog}~[3,-z_qe^{-\alpha}] - \mathrm{PolyLog}~[3,-z_qe^{\alpha}]\bigg).
\end{align}
Note that the results of Eqs.~\eqref{12}-\eqref{14} matches with that obtained in Ref.~\cite{Mitra:2016zdw}. We also observe that in the limit $z_q\to1$, Eqs.~\eqref{12}-\eqref{14} reduces to that obtained in Ref.~\cite{Jaiswal:2015mxa}.

The macroscopic definition of viscous tensor and the baryon diffusion current requires the non-equilibrium part of the distribution function of the particles. For the system close to local thermodynamic equilibrium, the non-equilibrium quasiparton phase space distribution function takes the form $f_k=f^0_k+\delta f_k$, where $\delta f_k/f^0_k \ll1$ and the equilibrium distribution $f^0_k$ are given in Eqs.~\eqref{2}-\eqref{4} for $k=(q,\bar{q},g)$. Macroscopically, the energy-momentum tensor in the non-equilibrium case can be decomposed as,
\begin{equation}\label{15}
T^{\mu\nu}=\varepsilon u^{\mu}u^{\nu} - (P+\Pi) \Delta^{\mu\nu} + \pi^{\mu\nu},
\end{equation}   
where $\Delta^{\mu\nu}\equiv g^{\mu\nu}-u^\mu u^\nu$ is the projection operator orthogonal to the fluid velocity, $\pi^{\mu\nu}$ is the shear stress tensor and $\Pi$ is the bulk viscous pressure. Similarly, the baryon four-current can be macroscopically described as,
\begin{equation}\label{16}
    N^{\mu}=n u^{\mu}+n^{\mu}.
\end{equation}
Note that the above expressions for energy-momentum tensor and particle four-current are written for fluid four-velocity defined in Landau frame.

The projection of $T^{\mu\nu}$ and $N^{\mu}$ conservation equations along and orthogonal to $u^{\mu}$ gives,
\begin{align}
    \dot{\varepsilon}+(\varepsilon+P+\Pi)\theta-\pi^{\mu\nu}\sigma_{\mu\nu}&=0,\label{17}\\
    (\varepsilon+P+\Pi)\dot{u}_{\alpha}-\nabla^{\alpha}(P+\Pi)+\Delta^\alpha_\nu\partial_{\mu}\pi^{\mu\nu}&=0,  \label{18}\\
    \dot{n}+n\theta+\partial_{\mu}n^{\mu}&=0   \label{19},
\end{align}
where $\theta\equiv\partial_{\mu}u^{\mu}$ is the expansion scalar, $\dot{A}\equiv u^\mu\partial_\mu A$ represents the comoving derivative, $\nabla^\alpha\equiv\Delta^{\alpha\beta}\partial_\beta$ is a space-like derivative operator which is orthogonal to $u^\alpha$ and $\sigma^{\mu\nu}\equiv\Delta^{\mu\nu}_{\alpha\beta}\nabla^{\alpha}u^{\beta}$. Here we define a four-index tensor $\Delta^{\mu\nu}_{\alpha\beta}\equiv\frac{1}{2}(\Delta^\mu_\alpha\Delta^\nu_\beta +\Delta^\mu_\beta\Delta^\nu_\alpha)-\frac{1}{3}\Delta^{\mu\nu}\Delta_{\alpha\beta}$ which is a traceless symmetric projection operator orthogonal to the fluid velocity. 

The expressions for the derivatives of $\alpha$ and $\beta$ can be obtained from Eqs.~\eqref{17}-\eqref{19} and takes the following form,
\begin{align}\label{20}
&\dot{\beta}=\chi_{\beta}\theta +\mathcal{O}(\delta^2), \quad \dot{\alpha}=\chi_\alpha\theta + \mathcal{O}(\delta^2),\\
&\nabla^{\mu}\beta=-\beta\dot{u}^{\mu}+\frac{n}{\varepsilon+P} \nabla^{\mu}\alpha+\mathcal{O}(\delta^2)\label{21},
\end{align}    
with $\chi_\beta$ and $\chi_\alpha$ taking the following form,
\begin{align}
&\chi_\beta = \Bigg[\frac{\tilde{J}_{q~10}^{(0)+} (\epsilon + P) - \tilde{J}_{q~20}^{(0) -}~n - \tilde{J}_{g~30}^{(0)}~n}{ \tilde{J}_{q~30}^{(0)+} \tilde{J}_{q~10}^{(0)+} - \tilde{J}_{q~20}^{(0)-} \tilde{J}_{q~20}^{(0)-} +  \tilde{J}_{g~30}^{(0)} \tilde{J}_{q~10}^{(0)+}}\Bigg], \label{22.0}\\
&\mathrm{and,} \nonumber\\
&\chi_\alpha = \Bigg[\frac{\tilde{J}_{q~20}^{(0)-} (\epsilon + P) - \tilde{J}_{q~30}^{(0) +}~n - \tilde{J}_{g~30}^{(0)}~n}{ \tilde{J}_{q~30}^{(0)+} \tilde{J}_{q~10}^{(0)+} - \tilde{J}_{q~20}^{(0)-} \tilde{J}_{q~20}^{(0)-} +  \tilde{J}_{g~30}^{(0)} \tilde{J}_{q~10}^{(0)+}}\Bigg]. \label{22}
\end{align}
Here $\tilde{J}_{k~nq}^{(r)\pm}$ are the thermodynamic integrals defined as
\begin{align}
\Tilde{J}^{(r)\pm}_{q~nm}&=\frac{g_q}{2\pi^2}\frac{(-1)^m}{(2m+1)!!}\int_{0}^\infty{d\mid\vec{\Tilde{p}}_q\mid}~\big(u.\Tilde{p}_q \big)^{n-2m-r-1}\nonumber\\
&\times\big(\mid\vec{\Tilde{p}}_q\mid\big)^{2m+2}f_q^\pm, \label{37}\\ 
    \Tilde{J}^{(r)}_{g~nm}&=\frac{g_g}{2\pi^2}\frac{(-1)^m}{(2m+1)!!}\int_{0}^\infty{d\mid\vec{\Tilde{p}}_g\mid}~\big(u.\Tilde{p}_g\big)^{n-2m-r-1}\nonumber\\
    &\times\big(\mid\vec{\Tilde{p}}_g\mid\big)^{2m+2}f_g\tilde{f}_g, \label{39}
\end{align}
where $f^\pm_q=f_q \tilde{f}_q \pm f_{\bar{q}} \tilde{f}_{\bar{q}}$ and $\tilde{f}_k\equiv(1- a{f}_k)$ with $a = -1$ and $+1$ for Bose-Einstein and Fermi-Dirac statistics, respectively. The expressions of these integral coefficients appearing in Eqs.~\eqref{22.0}~and~\eqref{22}, in terms of temperature and chemical potential, are given in Appendix~\ref{A}.

The shear stress tensor $\pi^{\mu\nu}$  can be expressed in terms of $\delta f_k$ within EQPM as follows~\cite{Mitra:2018akk},
\begin{align}\label{23}
\pi^{\mu\nu}=&\sum_k g_k\Delta^{\mu\nu}_{\alpha\beta}\int{d\Tilde{P}_k~ \Tilde{p}_k^{\alpha}\,\Tilde{p}_k^{\beta}\,\delta f_k}\nonumber\\
&+\sum_k g_k\,\delta \omega_k\, \Delta^{\mu\nu}_{\alpha\beta}\int{d\Tilde{P}_k~ \Tilde{p}_k^{\alpha}\,\Tilde{p}_k^{\beta}\,\frac{1}{E_k}\delta f_k},
\end{align}
where $k=(g, q,\bar{q})$ represents the particle species. Similarly, the bulk viscous pressure $\Pi$ and the particle diffusion current $n^{\mu}$ can also be defined as,
\begin{align}\label{24}
\Pi=&-\frac{1}{3}\sum_k g_k\Delta_{\alpha\beta}\int{d\Tilde{P}_k~    \Tilde{p}_k^{\alpha}\,\Tilde{p}_k^{\beta}\,\delta f_k}\nonumber\\
 &-\frac{1}{3}\sum_k g_k\,\delta \omega_k\,\Delta_{\alpha\beta}\int{d\Tilde{P}_k~ \Tilde{p}_k^{\alpha}\,\Tilde{p}_k^{\beta}\,\frac{1}{E_k}\delta f_k},
\end{align}
and 
\begin{align}\label{25}
n^{\mu}=&g_q\Delta_{\alpha}^{\mu}\int{d\Tilde{P}_q~\Tilde{p}_q^{\alpha}\, (\delta f_q-\delta f_{\bar{q}})}\nonumber\\
&-\delta\omega_q g_q\Delta_{\alpha}^{\mu}\int{d\Tilde{P}_q~ \Tilde{p}_q^{\alpha}\,\frac{1}{E_q}(\delta f_q-\delta f_{\bar{q}})}.
\end{align}
We will use the above equations for dissipative quantities to obtain their first-order expressions and corresponding transport coefficients.

The relativistic transport equation quantifies the rate of change of quasiparton phase space distribution function in terms of collision integral $C[f_k]$ and has the following form
\begin{equation}\label{26}
\Tilde{p}^{\mu}_k\,\partial_{\mu}f_k(x,\Tilde{p}_k)+F_k^{\mu}\left(u\!\cdot\!\tilde{p}_k\right)\partial^{(p)}_{\mu} f_k = C[f_{k}],
\end{equation}
where $F_k^{\mu}=-\partial_{\nu}(\delta\omega_k u^{\nu}u^{\mu})$ is the force term defined from the conservation of energy momentum and particle flow. In the current EQPM framework, the collision integral is defined in the relaxation time approximation (RTA), where the thermal relaxation $\tau_{R}$ linearizes the collision term as~\cite{Anderson_Witting}
\begin{equation}\label{27}
C[f_{k}]=-\left(u\!\cdot\!\tilde{p}_k\right)\frac{\delta f_k}{\tau_R}.
\end{equation}
To obtain $\delta f_k$, we solve the relativistic Boltzmann equation with RTA using the Chapman-Enskog like iterative expansion.

%%%%%%%%%%%%%%%%%%%%%%%%%%%%%%%%%%
 
\subsection{First order dissipative evolution equation}

The first order correction to distribution functions for quarks, anti-quarks and gluons can be obtained from the Boltzmann equation, Eq.~\eqref{26}, by considering an iterative Chapman-Enskog like solution \cite{Jaiswal:2013npa, Jaiswal:2013vta}. For the current effective kinetic theory, we obtain the following form,
\begin{align}
\!\!\delta f_q &= \tau_R\bigg[ \Tilde{p}_q^\gamma\partial_\gamma \beta \!+\! \frac{\Tilde{p}_q^\gamma}{u\!\cdot\!\Tilde{p}_q} \!\Big( \beta\, \Tilde{p}_q^\phi \partial_\gamma u_\phi \!-\! \partial_\gamma \alpha \Big) \!-\! \beta\theta\,\delta\omega_q \bigg]f_q\Tilde{f}_q, \label{28}\\
\!\!\delta f_{\bar{q}} &= \tau_R\bigg[ \Tilde{p}_{\bar{q}}^\gamma\partial_\gamma \beta \!+\! \frac{\Tilde{p}_{\bar{q}}^\gamma}{u\!\cdot\!\Tilde{p}_{\bar{q}}} \!\Big( \beta\, \Tilde{p}_{\bar{q}}^\phi \partial_\gamma u_\phi \!+\! \partial_\gamma \alpha \Big) \!-\! \beta\theta\,\delta\omega_{\bar{q}} \bigg] f_{\bar{q}} \Tilde{f}_{\bar{q}}, \label{29}\\
\!\!\delta f_g &= \tau_R\bigg( \Tilde{p}_g^\gamma\partial_\gamma \beta + \frac{\beta\, \Tilde{p}_g^\gamma\, \Tilde{p}_g^\phi}{u\!\cdot\!\Tilde{p}_g}\partial_\gamma u_\phi -\beta\theta\,\delta\omega_g \bigg) f_g\Tilde{f}_g, \label{30}
\end{align}
With the expressions for $\delta f_k $ obtained in Eqs.~\eqref{28}-\eqref{30}, we can obtain the first order evolution equation for dissipative quantities from Eqs.~\eqref{23}-\eqref{25}. 

Assuming the thermal relaxation time $\tau_{R}$ to be independent of particle four-momenta and keeping terms up to first-order in gradients, we obtain
\begin{align}
\pi^{\mu\nu} &= 2\,\tau_R\,\beta_\pi\,\sigma^{\mu\nu},\label{31}\\
\Pi &= -\tau_R\,\beta_\Pi\,\theta, \label{32}\\
n^\mu &= \tau_R\,\beta_n\,\nabla^\mu \alpha,\label{33}
\end{align}
where the coefficients have the following form,
\begin{align}
\beta_\pi=&~ \beta \bigg[\Tilde{J}^{(1)+}_{q~42}+\Tilde{J}^{(1)}_{g~42}+(\delta\omega_q)\Tilde{L}^{(1)+}_{q~42}+(\delta\omega_g)\Tilde{L}^{(1)}_{g~42}\bigg],\label{34}\\
\beta_\Pi=&~\beta\bigg[\frac{\chi_\beta}{\beta}\bigg(\Tilde{J}_{q~31}^{(0)+}+\Tilde{J}_{g~31}^{(0)}+(\delta\omega_q)\Tilde{L}_{q~31}^{(0)+}+(\delta\omega_g)\Tilde{L}_{g~31}^{(0)}\bigg)\nonumber\\ 
&+\frac{\chi_\alpha}{\beta}\bigg(\Tilde{J}_{q~31}^{(0)+}+\Tilde{J}_{g~31}^{(0)}+(\delta\omega_q)\Tilde{L}_{q~31}^{(0)+}+(\delta\omega_g)\Tilde{L}_{g~31}^{(0)}\bigg)\nonumber\\ 
&+\frac{5}{3}\bigg(\Tilde{J}_{q~42}^{(1)+}+\Tilde{J}_{g~42}^{(1)}+(\delta\omega_q)\Tilde{L}_{q~42}^{(1)+}+(\delta\omega_g)\Tilde{L}_{g~42}^{(1)}\bigg)\nonumber\\
&-(\delta\omega_q)\Tilde{J}_{q~21}^{(0)+} - (\delta\omega_g)\Tilde{J}_{g~21}^{(0)}\bigg],\label{35}\\
\beta_n=&~\bigg[\frac{n}{(\epsilon+P)}\bigg(\Tilde{J}^{(0)-}_{q~21}+\big(\delta\omega_q\big)\Tilde{L}^{(0)-}_{q~21}\bigg) - \Tilde{J}^{(1)+}_{q~21} \nonumber\\
    &- \big(\delta\omega_q\big)\Tilde{L}^{(1)+}_{q~21}\bigg].\label{36}
\end{align}
The thermodynamic integrals labeled by $\Tilde{L}^{(r)\pm}_{k~nm}$, appearing in the above expressions are defined as,
\begin{align}
\Tilde{L}^{(r)\pm}_{q~nm}&=\frac{g_q}{2\pi^2}\frac{(-1)^m}{(2m+1)!!}\int_{0}^\infty{d\mid\vec{\Tilde{p}}_q\mid}~\frac{\big(u.\Tilde{p}_q\big)^{n-2m-r-1}}{E_q}\nonumber\\
&\times\big(\mid\vec{\Tilde{p}}_q\mid\big)^{2m+2}f_q^\pm, \label{38}\\
\Tilde{L}^{(r)}_{g~nm}&=\frac{g_g}{2\pi^2}\frac{(-1)^m}{(2m+1)!!}\int_{0}^\infty{d\mid\vec{\Tilde{p}}_g\mid}~\frac{\big(u.\Tilde{p}_g\big)^{n-2m-r-1}}{\mid\vec{\Tilde{p}}_g\mid}\nonumber\\
    &\times\big(\mid\vec{\Tilde{p}}_g\mid\big)^{2m+2}f_g\tilde{f}_g. \label{40}
\end{align}
The expressions for the integral coefficients $\Tilde{J}^{(r)\pm}_{k~nm}$ and $\Tilde{L}^{(r)\pm}_{k~nm}$ appearing in Eqs.~\eqref{34}-\eqref{36} are given in Appendix~\ref{A} in terms of temperature and chemical potential.

By comparing the Eqs.~\eqref{31}-\eqref{33} with the relativistic Navier-Stokes equations~\cite{Landau},
\begin{align}\label{41}
    \pi^{\mu\nu}&=2\eta\,\sigma^{\mu\nu},
    &\Pi=-\zeta\,\theta,
    && n^\mu=\kappa_n\nabla^\mu\alpha,
\end{align}
we can obtain the coefficients of bulk viscosity, shear viscosity and charge conductivity as $\beta_{\pi}\tau_R=\eta$, $\beta_{\Pi}\tau_R=\zeta$ and $\beta_{n}\tau_R=\kappa_n$, respectively. Note that we consider a special case where the relaxation times for all particle species are same. The general case with different thermal relaxation time is left for future analysis. The form of above integrals for the massive and massless case, are presented in the Appendix~\ref{A}.

%%%%%%%%%%%%%%%%%%%%%%%%%%%%%%%%%%
\subsection{Viscous corrections to entropy-four current}

The entropy four-current $S^{\mu}$ has contributions from the shear viscous tensor and bulk viscous pressure for viscous fluids. One can derive the relativistic Navier-Stokes equations from local entropy generation, which is the divergence of entropy current $\partial_{\mu}S^{\mu}$, in the medium. The kinetic theory description of the entropy four-current from Boltzmann-H function takes the following form,
\begin{equation}\label{43}
    S^{\mu}=-\sum_{k}g_k\int{d\Tilde{P}_k}\Tilde{p}^{\mu}(f_k\ln{f_k}+a\tilde{f}_k\ln{\tilde{f}_k}).
\end{equation}
The viscous corrections contribute to the thermal distribution function as $f_k=f^0_k+\delta f_k$. The Eqs.~(\ref{28})-(\ref{30}) can be expressed in terms of first-order derivative of hydrodynamical variables by employing the shear and bulk viscous evolution equations and have the following forms,
\begin{equation}\label{44}
    \delta f_k=f_k^0{\tilde{f}}^0_k\phi_k,
\end{equation}
where $\phi_k=\phi_k^{\text{shear}}+\phi_k^{\text{bulk}}$ is the deviation from the equilibrium,
\begin{align}\label{45}
    &\phi_k^{\text{shear}}=\dfrac{\beta}{2\beta_{\pi}(u.\Tilde{p}_k)}\Tilde{p}_k^{\alpha}\Tilde{p}_k^{\beta}\pi_{\alpha\beta},\\
    &\phi_k^{\text{bulk}}=-\dfrac{\beta}{\beta_{\Pi}(u.\Tilde{p}_k)}\Big[(u.\Tilde{p}_k)^2\frac{\chi_\beta}{\beta}-\frac{\mid\vec{\Tilde{p}}_k\mid^2}{3}-(u.\Tilde{p}_k)\delta\omega_k\Big]\Pi.
\end{align}
We consider the case of vanishing chemical potential and at $\mu_q=0$, we have $\chi_\beta/\beta=c_s^2$, where $c_s^2$ is the square of the speed of sound in the medium. We obtain the viscous correction to the entropy current as,
\begin{align}\label{46}
    S^{\mu} &= s_0u^{\mu} - \int \mathrm{d\Tilde{P}}\, \Tilde{p}^\mu f_k^0\, \Tilde{f_k^0}\, \ln{\left(f_k^0/\Tilde{f}_k^0\right)}\, \phi_k + \mathcal{O}(\phi_k^2)+ .. ,
\end{align}
where $s_0=(\epsilon+P)/T$ is the equilibrium entropy density and the last term represents all the higher-order terms. Earlier studies have shown that first-order dissipative correction to the entropy current (the term proportional to $\phi_k$ in the above equation) vanishes and has non-zero contributions from the second and higher-order terms \cite{El:2009vj,Chattopadhyay:2014lya}. However, it is important to note that in the present case, the first-order correction is also non-vanishing due to mean field contributions.
The entire second-order corrections require the knowledge of distribution function away from equilibrium up to second-order derivatives of hydrodynamical variables. This is beyond the focus of the current analysis. In order to calculate the entropy four-current in terms of the hydrodynamic quantities up to first order, we substitute Eq.~(\ref{44}) in Eq.~(\ref{46}). After performing the thermodynamic integrations, we obtain 
\begin{align}\label{47}
    S^{\mu}&=s_0u^{\mu} -  \frac{\beta}{\beta_\Pi} u^\mu\, \lambda_{\Pi}\, \Pi, 
\end{align}
where 
\begin{align}\label{coefficients of entropy correction}
    \lambda_{\Pi} &= \sum_k\Bigg[\ln{z_k} \bigg(\frac{\chi_\beta}{\beta} \Tilde{J}^{(0)}_{k,\, 20} + \Tilde{J}^{(0)}_{k,\, 21} - \delta\omega_k\, \Tilde{J}^{(0)}_{k,\, 10}\bigg)\nonumber\\
    &- \beta \bigg(\frac{\chi_\beta}{\beta}\, \Tilde{J}^{(0)}_{k,\, 30} + \Tilde{J}^{(0)}_{k,\, 31} - \delta\omega_k\, \Tilde{J}^{(0)}_{k,\, 20}\bigg)\Bigg].
\end{align}
The details of the thermodynamic integrals used here are given in Appendix \ref{A}.

Note that the first-order viscous correction is solely from the bulk viscous part. As a consequence of the Landau frame condition $u_\mu \pi^{\mu\nu} = 0$, the tensor structure of $S^\mu$ does not allow any contribution from the shear stress tensor. Note that in the ultra-relativistic limit, i.e., when mean field contributions can be ignored, the coefficient of first-order bulk viscous corrections to the entropy density, $\lambda_\Pi$, vanishes which is consistent with the results of previous studies~\cite{El:2009vj,Chattopadhyay:2014lya}. This implies that the non-vanishing first-order correction is due to the quasiparticle excitation in the medium. The effect of bulk viscous corrections to the entropy current can be quantified for the case of longitudinal Bjorken expansion. 

%---------------------------------
\begin{figure*}
\centering
\subfloat{\includegraphics[scale=0.4]{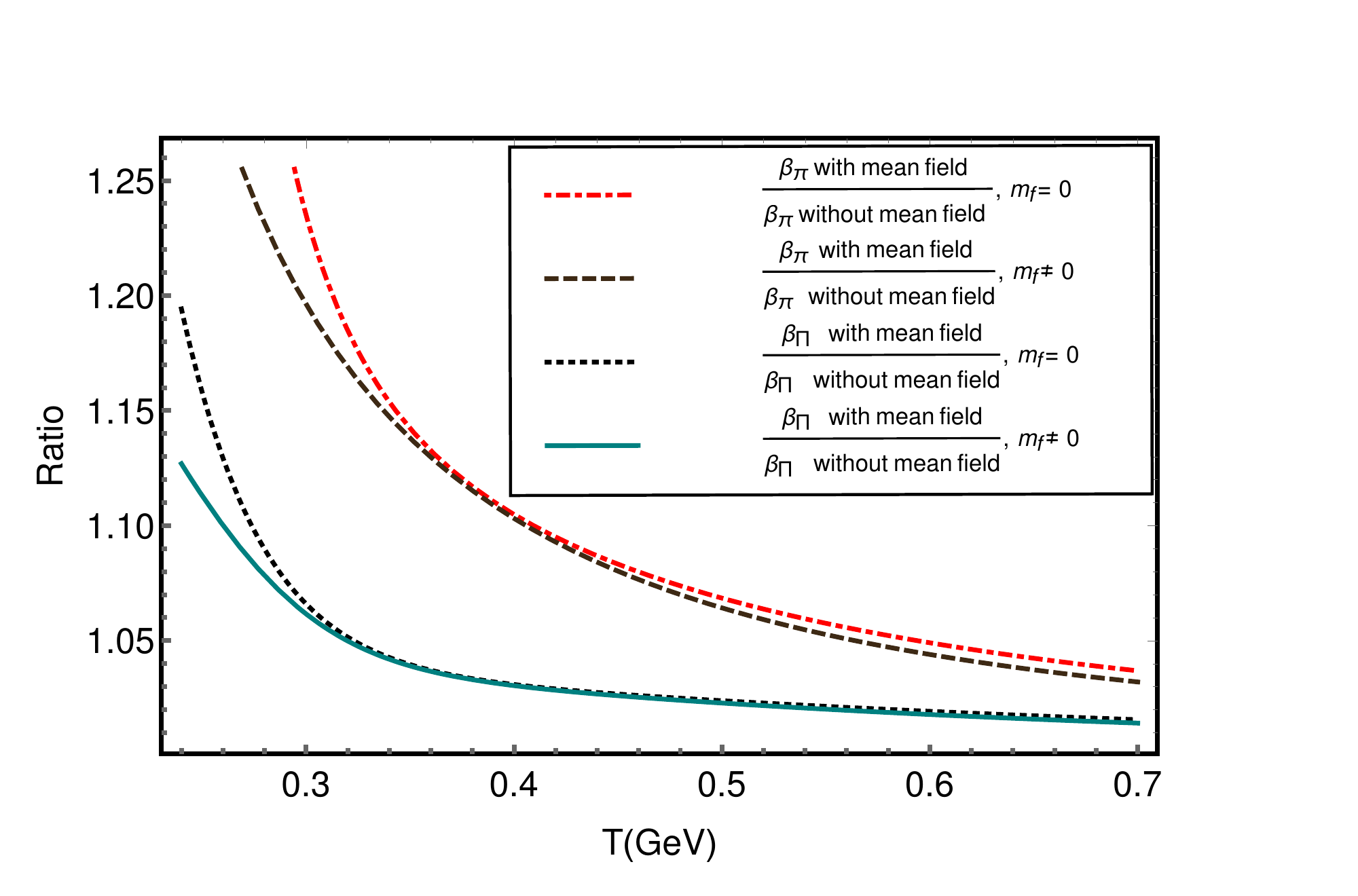}}
\hspace{.5 cm}
\subfloat{\includegraphics[scale=0.45]{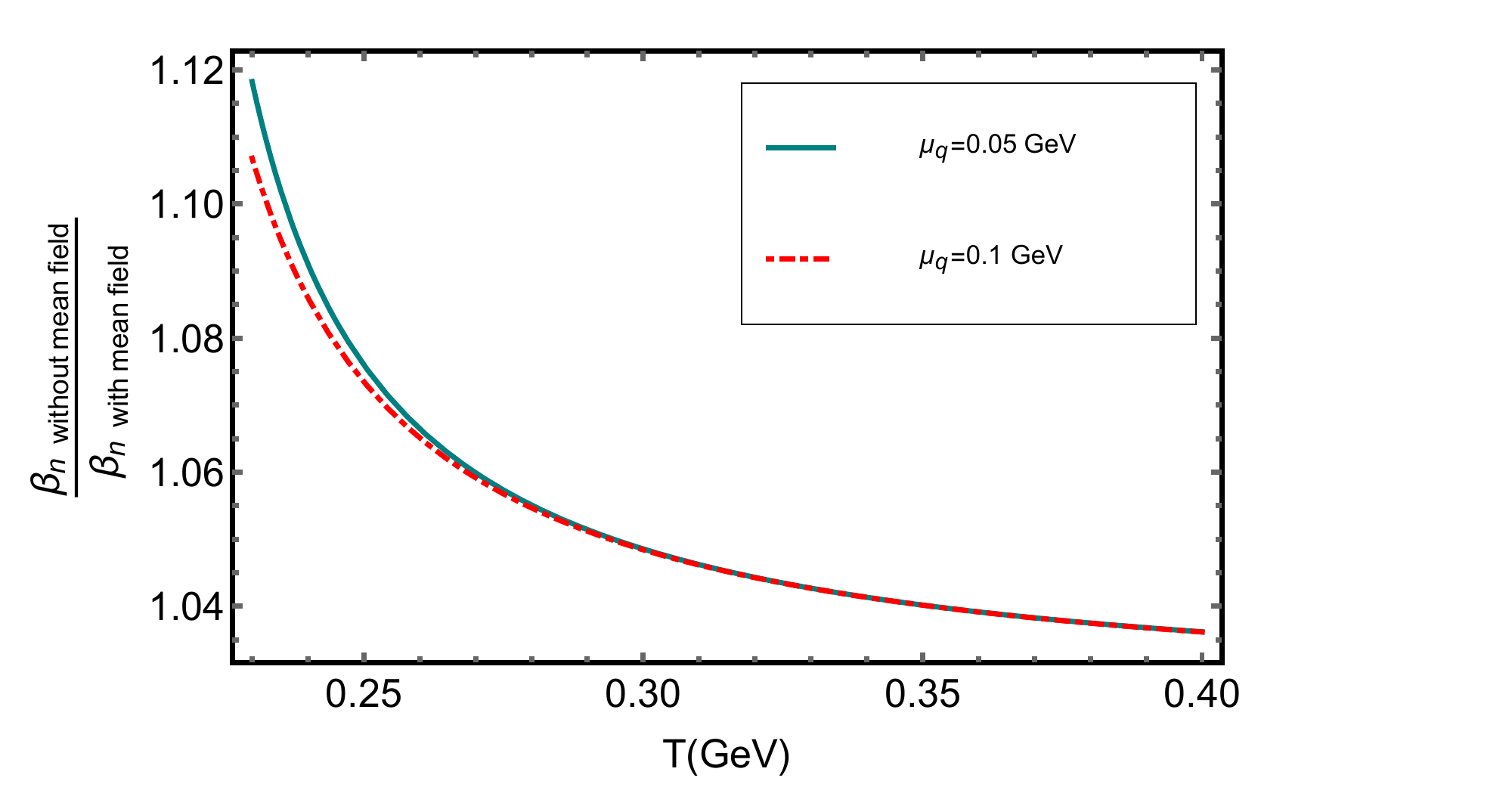}}
\caption{(Left panel) The effect of mean field contributions to the coefficients of bulk viscous pressure, shear tensor at $\mu=0.1$  GeV with and without quark mass correction. (Right panel) The mean field contribution to the particle diffusion at different quark chemical potential as a function of temperature.}
\label{f1}
\end{figure*}
%---------------------------------
%%%%%%%%%%%%%%%%%%%%%%%%%%%%%%%%%%%%%%%%%%%%%%%%%%%%%%%%%%%%%%%%%%%%%%

\subsection{Longitudinal boost-invariant expansion}

To model the dissipative hydrodynamical evolution of the QGP formed in the heavy-ion collision experiments, we employ the Bjorken's prescription~\cite{Bjorken:1982qr} for one-dimensional boost invariant expansion. Here, we consider the case of vanishing baryon chemical potential. The evolution equation of the energy density for purely longitudinal boost-invariant expansion can be expressed in terms of Milne coordinates $(\tau,x,y,\eta_s)$, where $\tau=\sqrt{t^2-z^2}$ and $\eta_s=\tanh^{-1}(z/t)$ resulting in $u^{\mu}=(1,0,0,0)$ with the metric tensor given by $g^{\mu\nu}=(1,-1,-1,-1/\tau^2)$~\cite{Tinti:2016bav}. Employing the Milne coordinate system, the energy evolution equation Eq.~(\ref{17}) gets simplified to,
\begin{equation}\label{42}
\frac{d\varepsilon}{d\tau}=-\bigg(\frac{\varepsilon+P}{\tau}\bigg)+\bigg(\frac{\zeta+4\eta/3}{\tau^2}\bigg),
\end{equation}
where we have used $\theta = 1/\tau,$ $\Pi=-\zeta/\tau$, $\pi^{\mu\nu}\sigma_{\mu\nu} = \Phi/\tau$ and $\Phi=4\eta/3\tau$. We numerically solve Eq.~(\ref{42}) to study the evolution of viscous nuclear matter with the values of dissipative quantities given in Eq.~(\ref{34}) and  Eq.~(\ref{35}), imposing the LEoS. The initial condition in RHIC (for Pb-Pb collision) is $T = 0.36$~GeV at $\tau_0 = 0.6$~fm/c and in LHC (for Au-Au collision) is $T = 0.5$~GeV at $\tau_0 = 0.4$~fm/c \cite{El:2007vg}. We estimated the temperature evolution by assuming relaxation time to be same for both bulk and shear parts ($\tau_R = \tau_\pi = \tau_\Pi = 0.25$~fm/c). With these conditions, we can investigate the proper time evolution of longitudinal pressure ($P_L$) to transverse pressure ($P_T$), $P_L/P_T\equiv (P+\Pi-\Phi)/(P+\Pi+\Phi/2)$, where $P$ is the equilibrium thermodynamic pressure.

%----------------------------------
\begin{figure}[t]
 \subfloat{\includegraphics[scale=0.35]{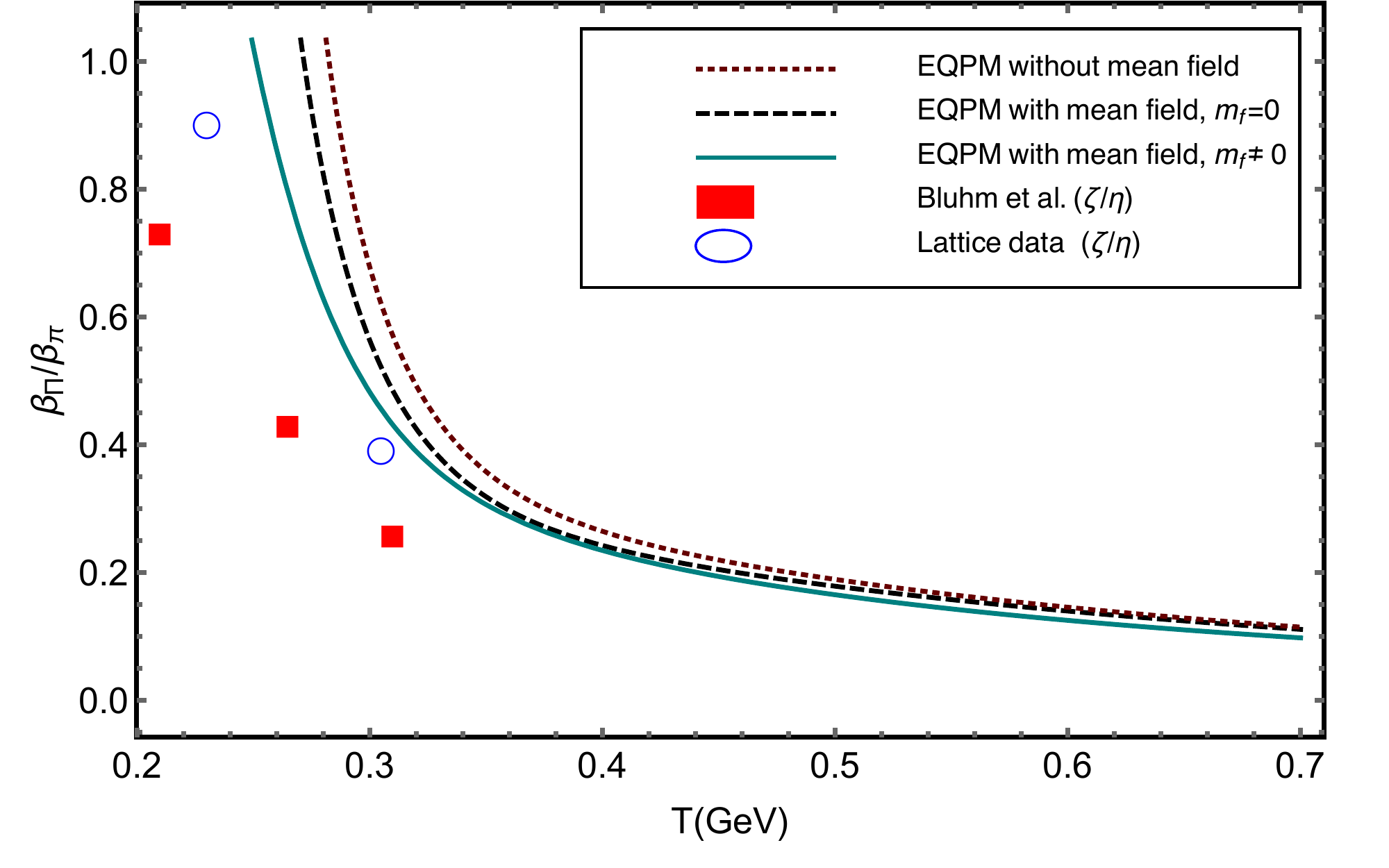}}
\caption{(Color online) The ratio $\frac{\beta_{\Pi}}{\beta_{\pi}}$ as a function of temperature  and comparison with the results in~\cite{Bluhm:2011xu, Meyer:2007ic,Meyer:2007dy}. }
\label{f2}
\end{figure} 
%---------------------------------
%%%%%%%%%%%%%%%%%%%%%%%%%%%%%%%%%%%%%%%%%%%%%%%%%%%%%%%%%%%%%%%%%%%%%%%%

\section{Results and discussions}

%---------------------------------
\begin{figure*}
 \centering
 \subfloat{\includegraphics[scale=0.45]{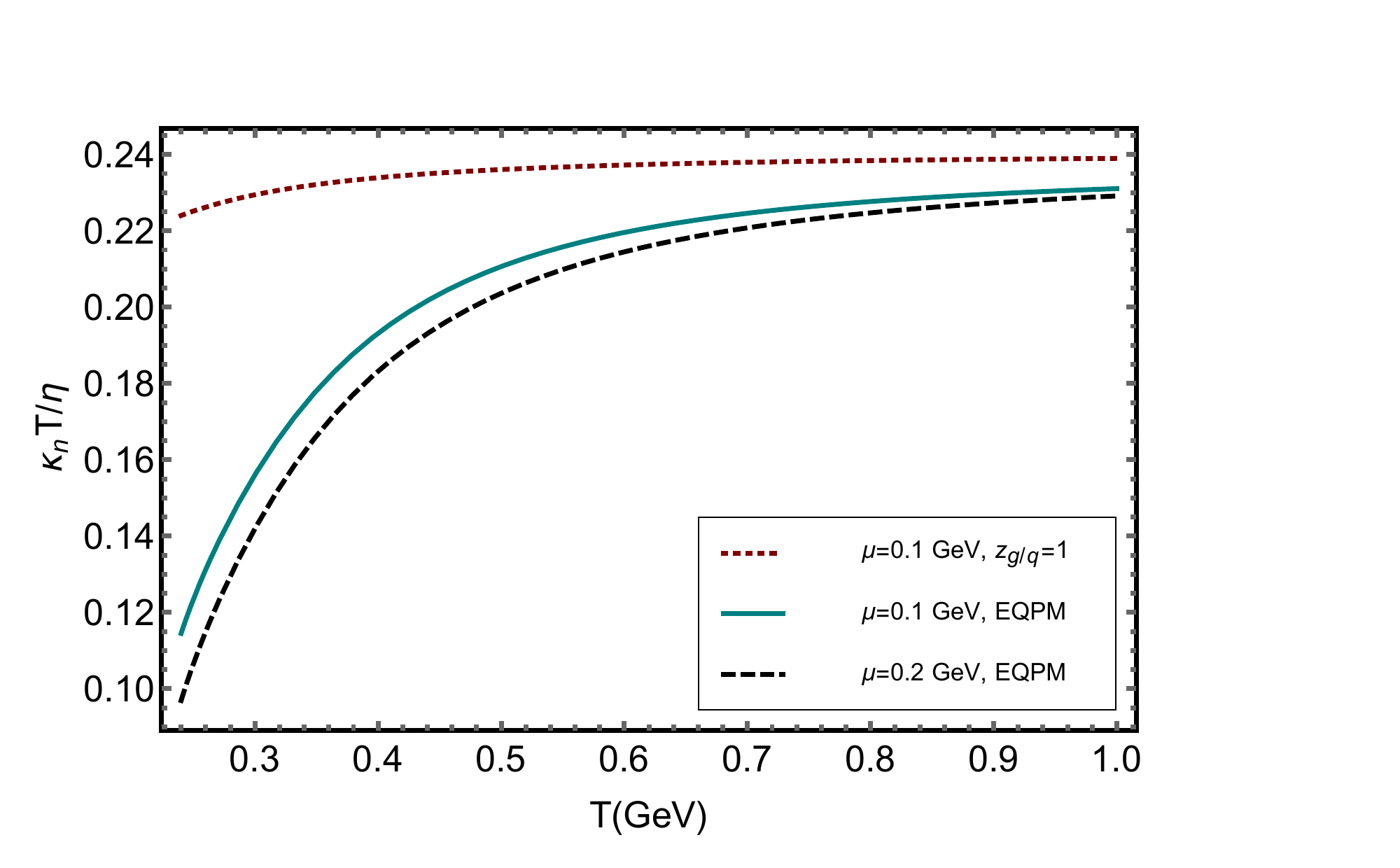}}
% \hspace{-.05 mm}
 \subfloat{\includegraphics[scale=0.45]{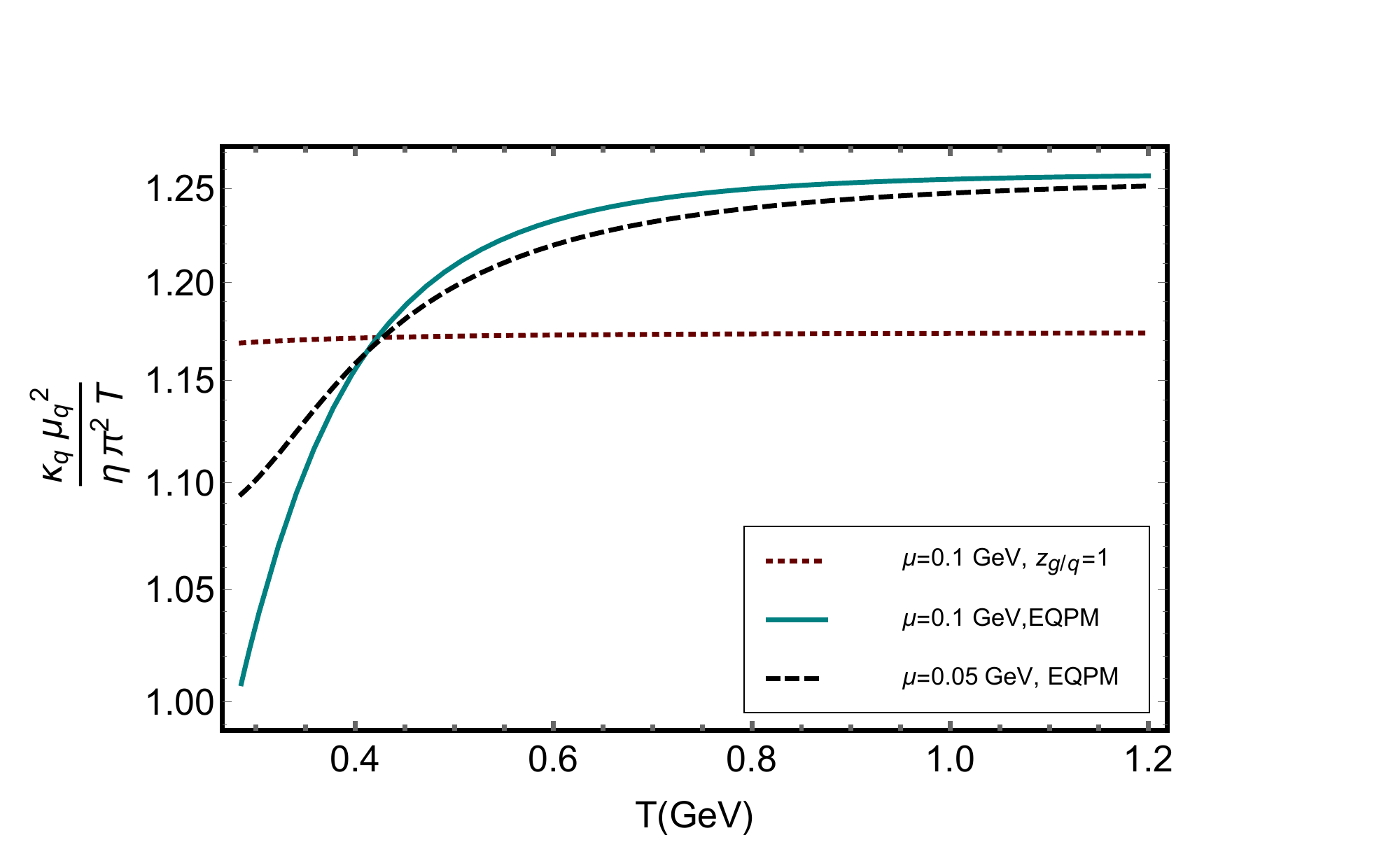}}
\caption{(Color online) The temperature behavior of the ratio $\frac{\kappa_nT}{\eta}$ (left panel) and $\frac{\kappa_q\mu^2_q}{\eta\pi^2T}$ (right panel) for different quark chemical potential.}
\label{f3}
\end{figure*} 
%---------------------------------
\begin{figure*}
 \centering
% \hspace{-.5 cm}
 \subfloat{\includegraphics[scale=0.44]{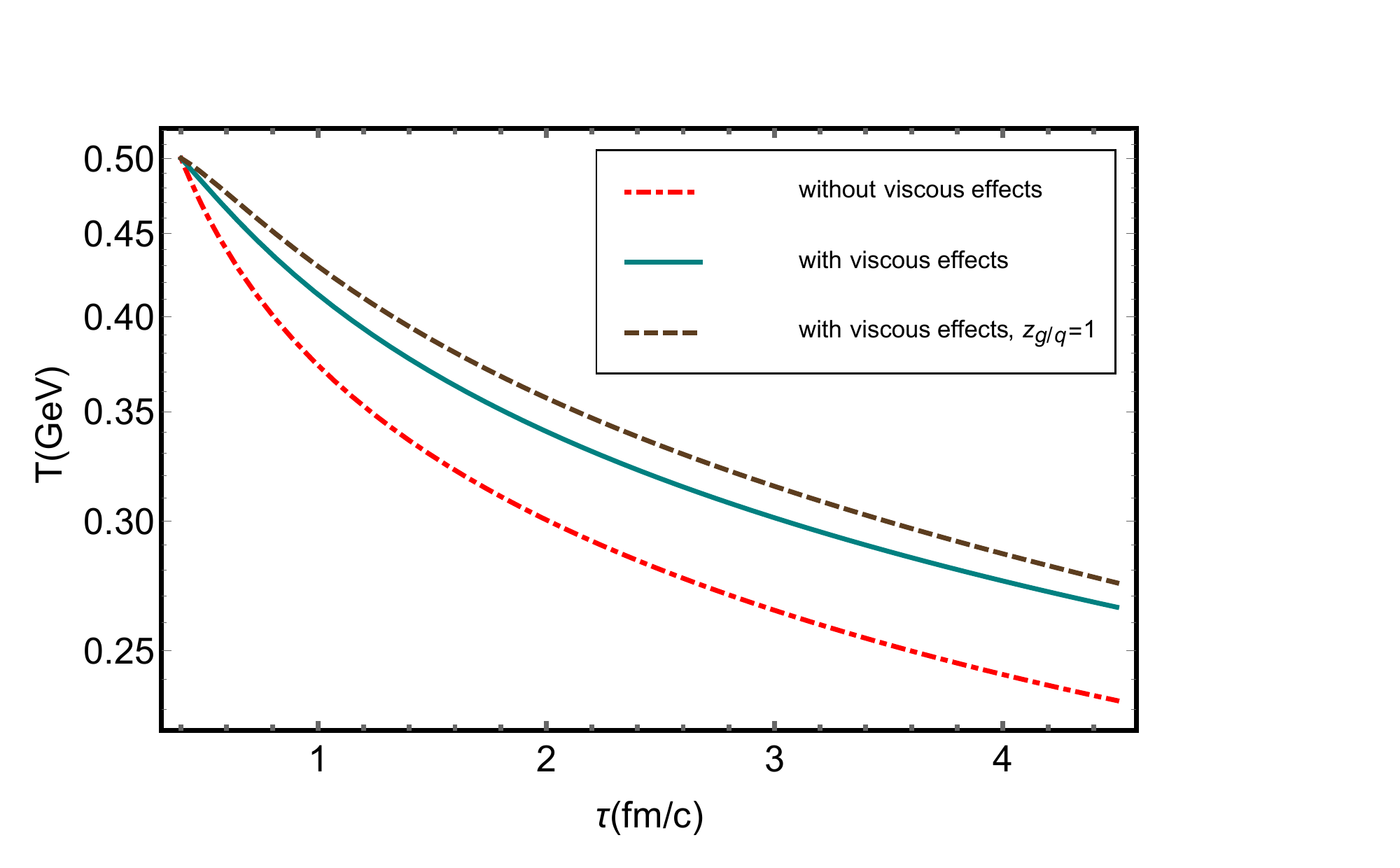}}
 \hspace{.6 cm}
 \subfloat{\includegraphics[scale=0.35]{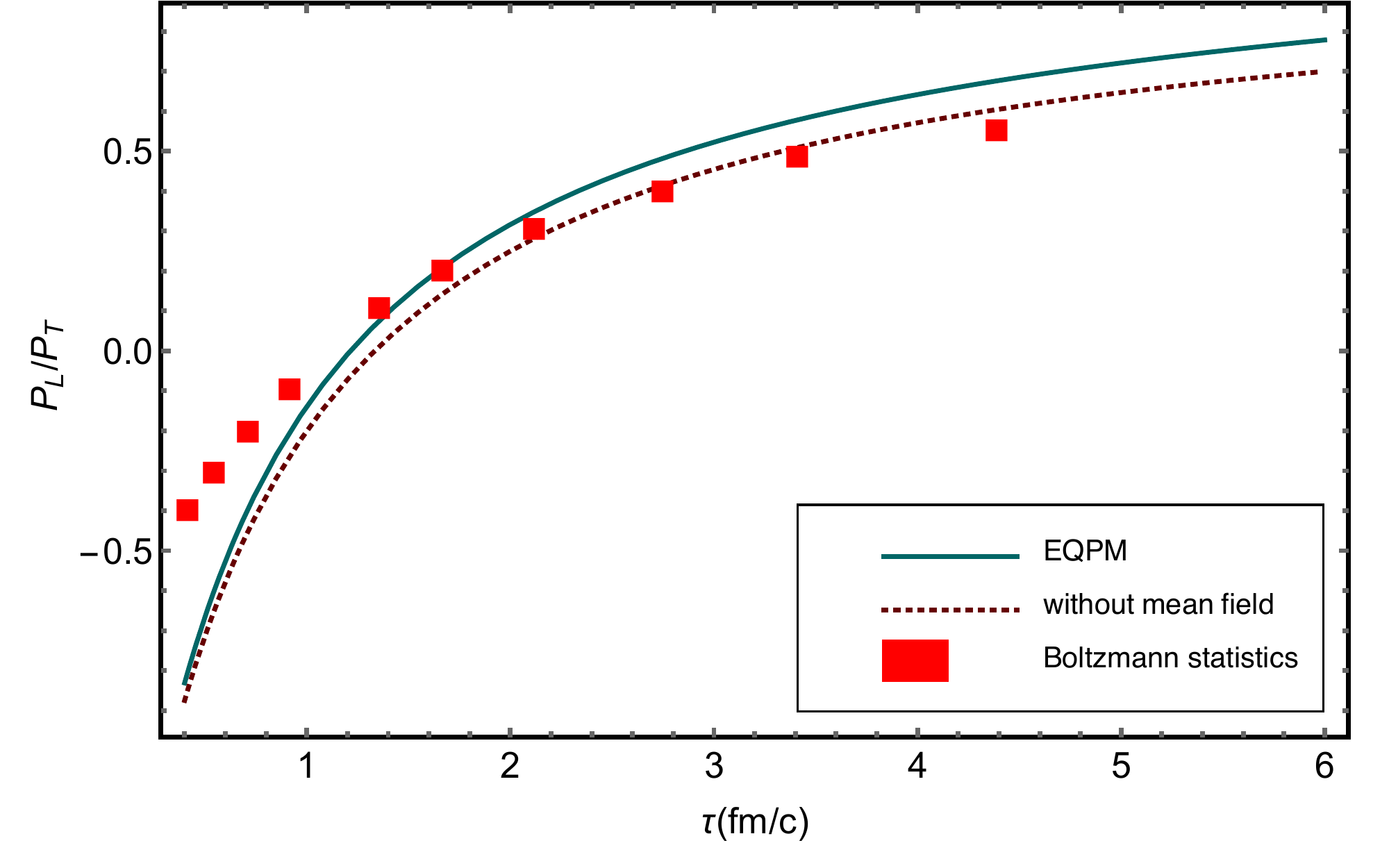}}
% \hspace{1.7 cm}
\caption{(Color online) The proper time evolution of  temperature (left panel) and pressure anisotropy (right panel) with initial temperature $T_0=500$~MeV at proper time $\tau_0=0.4$~fm/c. The behavior of $P_L/P_T$ is compared with the result obtained using higher-order corrections to Boltzmann statistics \cite{Jaiswal:2013vta}.}
\label{f4}
\end{figure*}   
%--------------------------------- 
%----------------------------------
\begin{figure*}
 \centering
 \hspace{-2.5 cm}
 \subfloat{\includegraphics[scale=0.35]{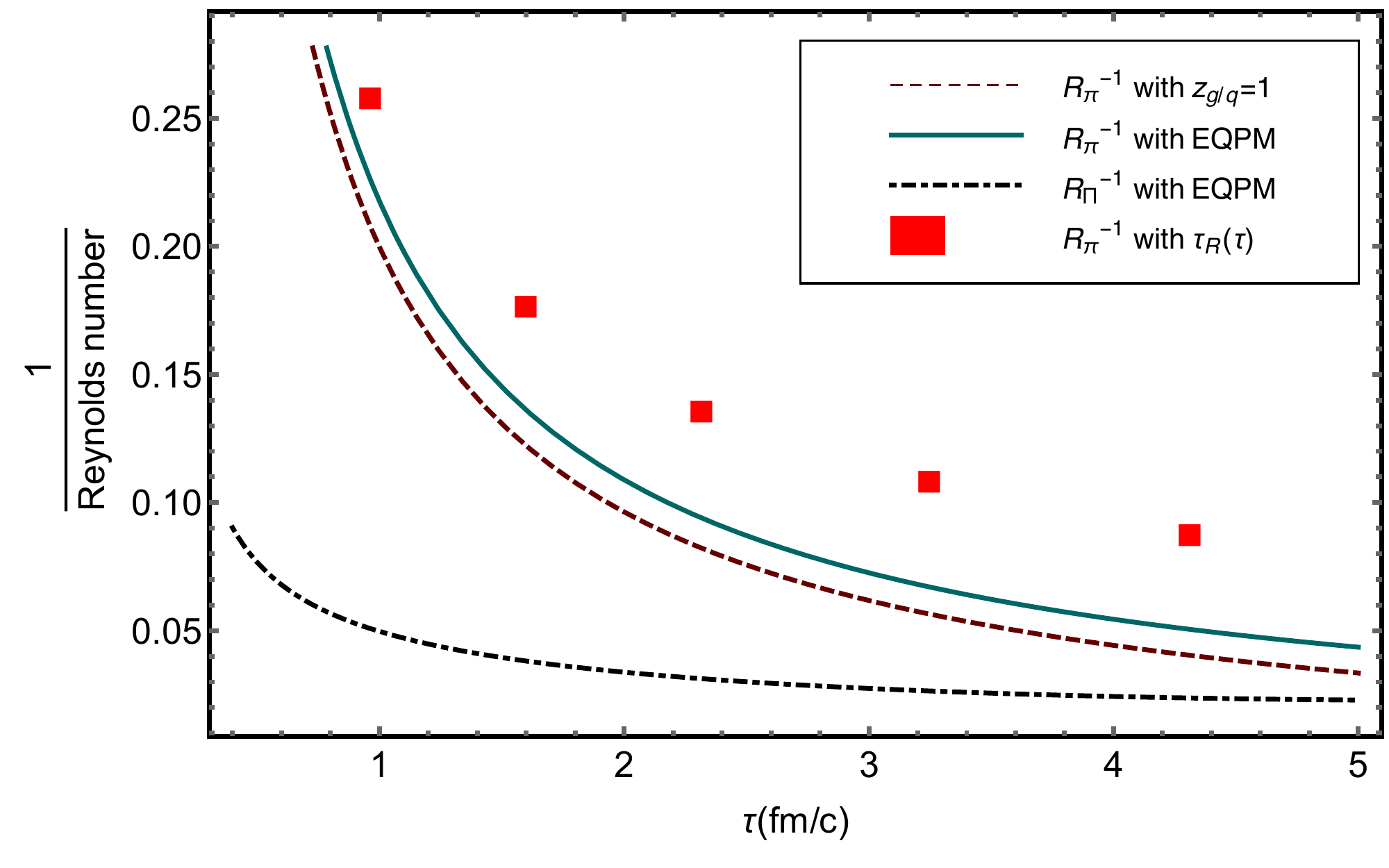}}
 \hspace{2.2 cm}
 \subfloat{\includegraphics[scale=0.35]{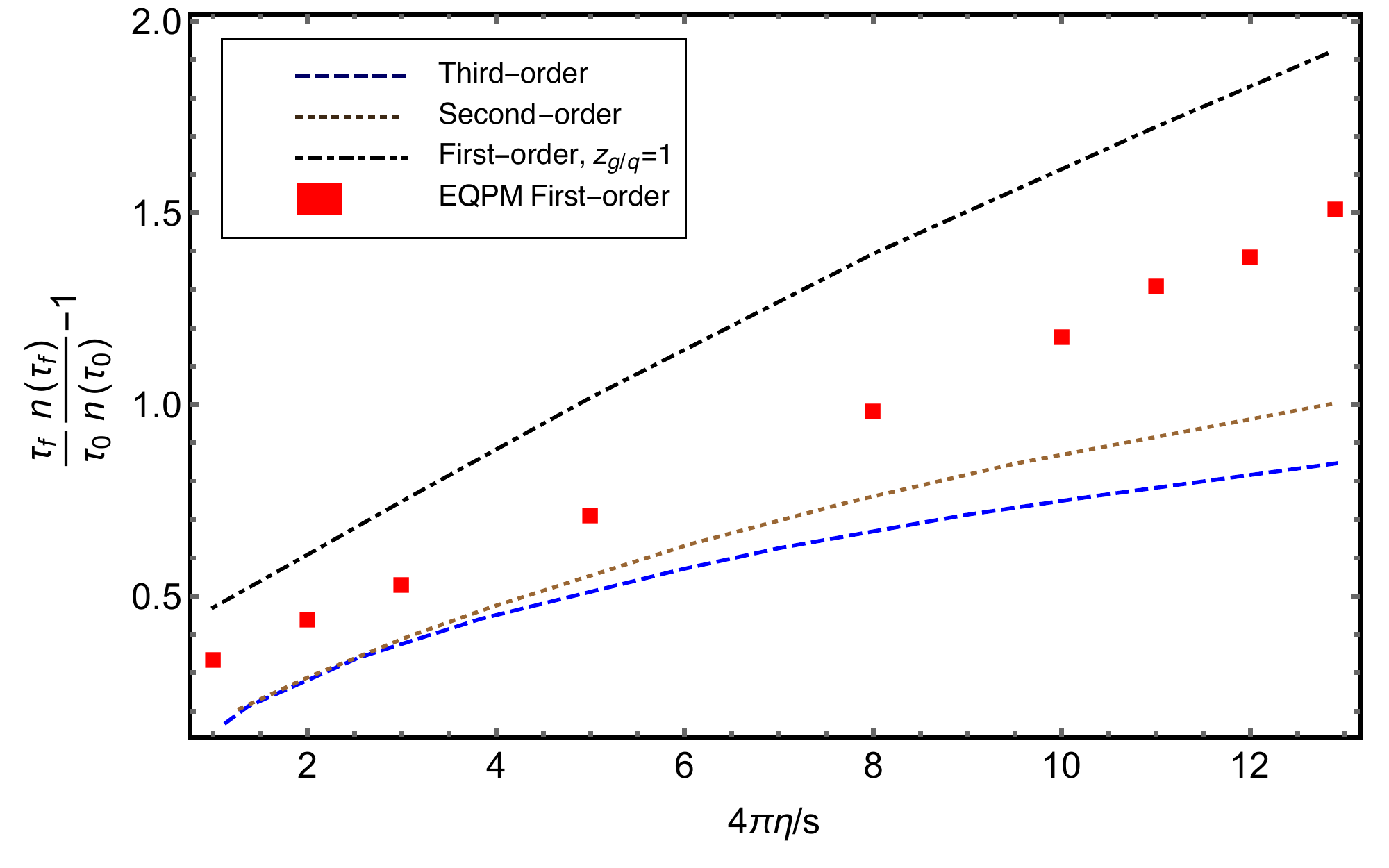}}
\caption{(Color online) Proper time evolution of Inverse of Reynolds number (left panel). The particle production measure as a function of $4\pi\eta/s$ (right panel). }
\label{f44}
\end{figure*} 
%---------------------------------
%
\begin{figure*}
% \centering
% \hspace{-.5 cm}
 \subfloat{\includegraphics[scale=0.415]{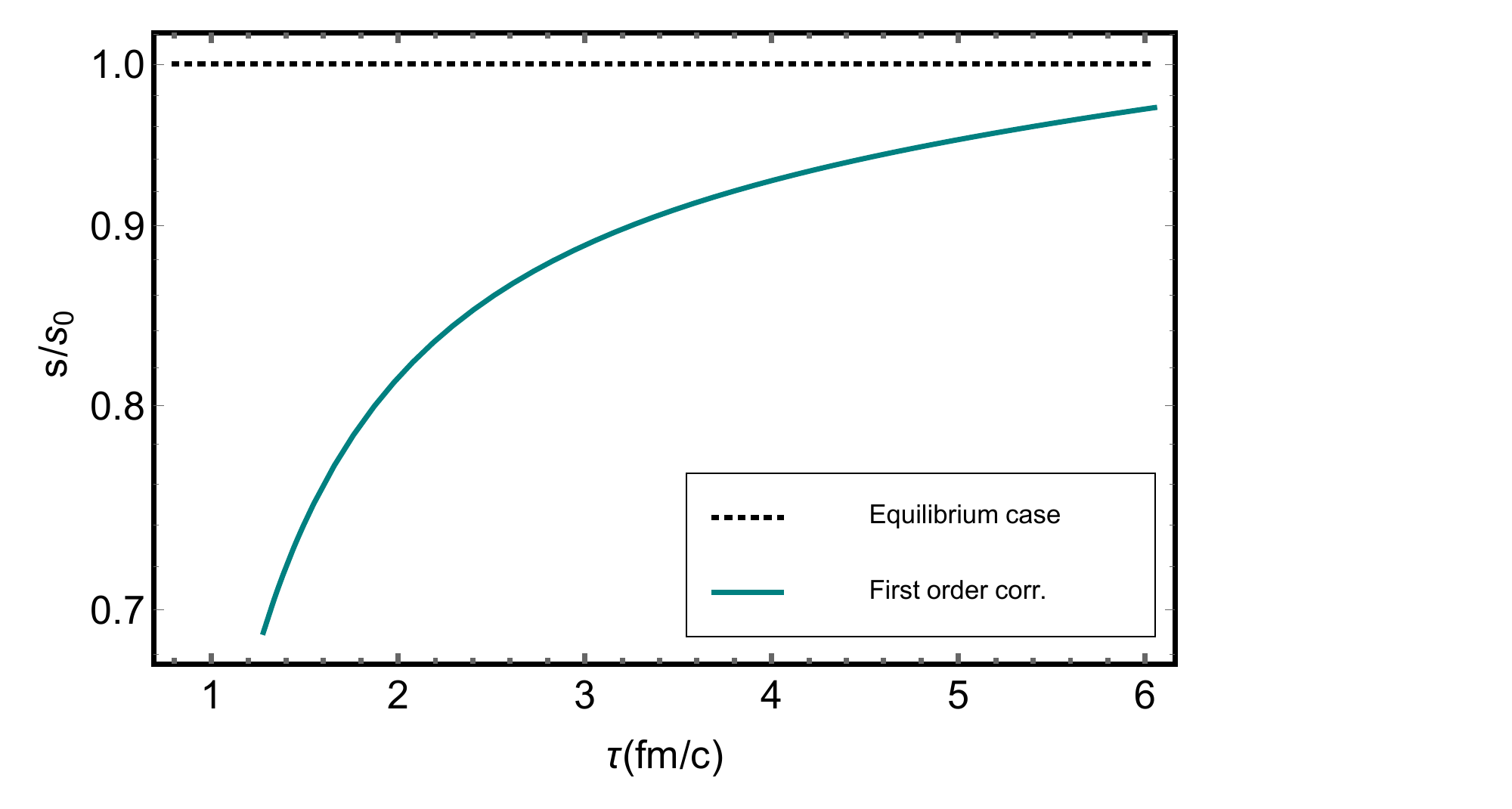}}
 \hspace{.8 cm}
 \subfloat{\includegraphics[scale=0.415]{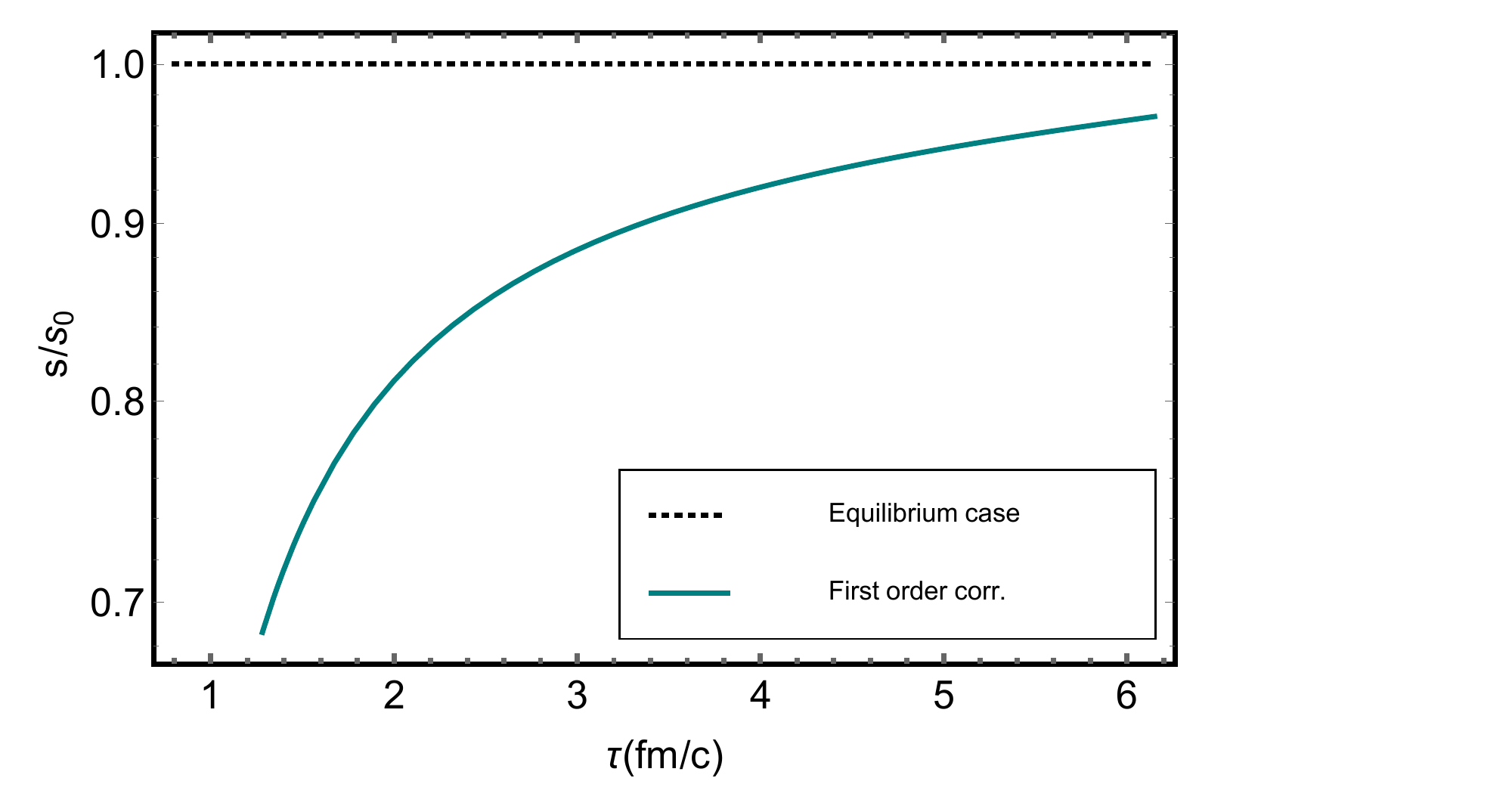}}
\caption{(Color online) First order viscous effects to the proper time evolution of $s/s_0$ for initial temperature $T_0=360$~MeV at proper time $\tau_0=0.6$~fm/c (left panel) and $T_0=500$~MeV at $\tau_0=0.4$~fm/c (right panel).}
\label{f5}
\end{figure*}   
%--------------------------------- 
\subsection{Mean-field effects and relative significance of transport coefficients}

We initiate the discussion with the temperature dependence of mean field corrections to the shear tensor, bulk viscous pressure, and the particle diffusion, respectively, of the hot QGP with finite quark chemical potential.
The dependence of finite quark mass and baryon chemical potential to the mean field contributions are separately shown in the left and right panel of Fig.~\ref{f1}, respectively. The mean field correction to the transport parameters with binary, elastic collisions at $m_q=0$ and $\mu_q=0$ is described in Ref.~\cite{Mitra:2018akk}. The effects of quark mass and chemical potential are visible in the low temperature regimes, whereas in the higher temperature regimes, the mean field contributions are almost independent on $m_q$ and $\mu_q$.
In Fig.~\ref{f2}, we show the temperature dependence of the ratio of the coefficient of the bulk viscous tensor to that of the shear tensor $\frac{\beta_{\Pi}}{\beta_{\pi}}$ at $\mu_q=0.1$ GeV. In the RTA, the ratio becomes $\frac{\beta_{\Pi}}{\beta_{\pi}}=\frac{\zeta}{\eta}$, where  $\zeta$ is the bulk viscosity of the hot QGP medium. 
We observe that the quark mass correction and mean field corrections are more visible in the low temperature regime near to the transition temperature $T_c$. Further, we compared the results with other parallel work and the lattice results. We found that our observations are consistent with the results of~\cite{Bluhm:2011xu, Meyer:2007ic, Meyer:2007dy}. 
The relative significance of charge conductivity $\kappa_n$ and shear viscosity $\eta$ could be understood in terms of the ratio $\kappa_nT/\eta$. Within RTA, the quantity $\beta_{n}T/\beta_{\pi}= \kappa_nT/\eta$. The temperature behavior of the ratio  $\kappa_nT/\eta$ is plotted in Fig.~\ref{f3} (left panel) for different quark chemical potential. The ratio becomes almost constant at high temperature regions, whereas it drops for the low temperature regime, indicating that the conductivity of the medium is relatively small to the shear viscosity in the regime $T\leq 2.5 T_c$ as compared to very high temperature regimes. We observe a similar trend in the temperature behavior of the dimensionless ratio $\frac{\kappa_q\mu^2_q}{\eta\pi^2T}$, in which $\kappa_q$ is the coefficient of thermal conductivity. The quantity $\frac{\kappa_q}{\eta}$ is defined as $\frac{\kappa_q}{\eta}=(\beta_{n}/\beta_{\pi})(\frac{\varepsilon+P}{nT})$ within RTA~\cite{Jaiswal:2015mxa}. In the high temperature limit, the ratio reduces to $\frac{\kappa_q}{\eta}=C\frac{\pi^2T}{\mu_q^2}$ as shown in the Fig.~\ref{f3} (right panel). 
For the non-interacting QGP $z_k=1$, the value of the constant becomes $C=95/81$~\cite{Jaiswal:2015mxa}. However, it should be noted that for the realistic EoS, the value of $C$ is equal to $5/4$.

%%%%%%%%%%%%%%%%%%%%%%%%%%%%%%%%%%%%%%%%%%%%%%%%%%%%%%

\subsection{Temperature evolution and pressure anisotropy}

In Fig.~\ref{f4} (left panel), we show the proper time evolution of temperature and pressure anisotropy in ideal and first order hydrodynamics with initial temperature $T_0=500$ MeV at proper time $\tau_0=0.4$ fm/c. We also plot temperature evolution with viscosity but without mean field effects, i.e., $z_{g/q}=1$. We assumed the Navier-Stokes initial condition for shear and bulk viscous part, respectively, as $\Phi=4\eta/3\tau_0$ and $\Pi=-\zeta/\tau_0$, with the thermal relaxation time $\tau_R=0.25$ fm/c. The temperature evolution based on the first order dissipative hydrodynamics shows a slower temperature drop with proper time compared to the ideal evolution. However, we also see that the mean field effects lead to faster cooling compared to the case when mean fields are not considered. This may be attributed to the fact that the mean filed takes away part of the internal energy during the evolution.

Following the temperature evolution, the proper time dependence of the pressure anisotropy $P_L/P_T$ is shown in Fig.~\ref{f4} (right panel). We see that there is a slightly faster approach to isotropization in the present EQPM model compared to the case when mean fields are absent. We also compare our results with those obtained using higher-order corrections to Boltzmann statistics \cite{Jaiswal:2013vta} and find faster isotropization in the present EQPM model. We note that the ratio $P_L/P_T$ approaches to negative value around $\tau=1$ fm/c and this implies that the bulk viscous pressure and pressure due to the shear tensor is larger than the absolute value of the thermodynamic pressure in the early stages of collision. Negative values of $P_L/P_T$ indicate mechanical instabilities in the system, which may lead to cavitation in the medium, indicating the breakdown of hydrodynamical expansion. This is, however, not surprising as the dissipative pressures due to first-order Navier-Stokes equation are quite large at early times leading to negative longitudinal pressure. Negative longitudinal pressure may also arise in higher-order hydrodynamic evolution due to the choice of initial conditions and transport coefficients \cite{Strickland:2014pga, Torrieri:2008ip}. 

%%%%%%%%%%%%%%%%%%%%%%%%%%%%%%%%%%%%%%%%%%%%%%%%%%%%%%%%%%%%%%%%%

\subsection{Reynolds number and particle production measure}

The system response followed by the viscous flow can be characterized by Reynolds number and has the following form~\cite{Jeon:2015dfa, Denicol:2012cn, Muronga:2003ta}, 
\begin{equation}\label{Reynolds pi and Pi}
    R^{-1}_\pi=  \frac{\sqrt{\pi_{\mu\nu}\pi^{\mu\nu}}}{P}~, \qquad R^{-1}_\Pi=  \frac{\lvert\,\Pi\,\rvert}{P},
\end{equation}
where $R^{-1}_{\pi}$ and $R^{-1}_{\Pi}$ denotes the inverse of Reynolds numbers associated with shear and bulk viscous pressure, respectively. In the case of a perfect fluid, the inverse Reynolds numbers vanish.
In Fig.~\ref{f44} (left panel), we show the proper time evolution of the inverse Reynolds number associated with shear and bulk viscous pressure. Here, we consider the case of vanishing quark mass and zero chemical potential. The EoS dependence of $R^{-1}_\pi$ is more pronounced in the later stage of expansion. Note that the finite contribution of $R^{-1}_\Pi$ is solely from the mean-field corrections, and vanishes for a massless system at $z_{g/q}=1$. We observe that the contribution from shear-stress tensor is dominant over the bulk viscous pressure in the medium. We compare our result with that obtained in Ref.~\cite{Muronga:2003ta} for the shear case, where the authors have employed a particular choice of thermal relaxation time.  

The particle production measure $\Delta_n$ is another interesting quantity to study the phenomenological effect of viscous quantities in the medium. Since the proper-time, $\tau$, controls fireball volume for a Bjorken expanding system, the particle production measure is given by~\cite{Strickland:2014pga, Bazow:2013ifa},
\begin{equation}\label{ppm}
\Delta_n = \frac{\tau_f\, n(\tau_f)}{\tau_0\, n(\tau_0)} - 1,
\end{equation}
where $\tau_f$ is the chemical freeze-out time. Here, $n(\tau_0)$ and $n(\tau_f)$ denotes the number density at $\tau_0$ and $\tau_f$, respectively. The quantity $\Delta_n$ vanishes for non-dissipative evolution and therefore it measures the particle production in the medium due to entropy production and viscous heating.
We plot the particle production measure as a function of $4\pi\eta/s$ in the case of first order viscous hydrodynamic evolution in  Fig.~\ref{f44} (right panel). We take the initial proper time, $\tau_0 = 0.4$~fm/c with $T(\tau_0)=500$ MeV. The freeze-out proper time, $\tau_f$, can be obtained from the temperature evolution under Bjorken flow by setting the chemical freeze-out temperature $T_f$. The temperature range of chemical freeze-out is explored in Refs.~\cite{BraunMunzinger:2003zd, Cleymans:2005xv, Biswas:2020dsc, Bhattacharyya:2019cer} and we consider $T_f = 170$~Mev in the current analysis. The EoS has a significant effect on the $\Delta_n$, and we observe a suppression in the EQPM description of particle production as compared to a first-order viscous formulation with ideal EoS. This suppression in particle production should also be reflected in the viscous entropy correction, as can be seen in Sec. III-C results. 

 In Fig.~\ref{f44} (right panel), we see that the particle production measure vanishes asymptotically in the limit of ideal hydrodynamics $\eta/s\rightarrow 0$ (entropy conservation). We compare our results with those obtained from higher-order viscous hydrodynamic evolution~\cite{Bazow:2013ifa}. We also note that traditional formulations of dissipative hydrodynamics do not capture the free streaming limit ($\tau_R\to\infty$ or equivalently $\eta/s\to\infty$) of the Boltzmann equations and therefore $\Delta_n$ does not vanish for large viscosities. This is due to the fact that traditional hydrodynamic formulations assume small departure from equilibrium, which is not satisfied for large viscosities. On the other hand, it is worth mentioning that the free streaming limit is correctly captured in the formulation of anisotropic hydrodynamics, which has no such restrictions; see Ref.~ \cite{Strickland:2014pga} for a pedagogical review. 

%%%%%%%%%%%%%%%%%%%%%%%%%%%%%%%%%%%%%%%%%%%%%%%%%%%%

\subsection{Entropy four-current in viscous medium}

In Sec.~II-C, we obtained expressions for the dissipative correction to entropy four-current from where one can obtain the non-equilibrium entropy density $s=u_\mu S^\mu$. The time evolution of the entropy density in the viscous medium, scaled by its equilibrium value, $s/s_0$, is plotted in Fig.~\ref{f5} with  dissipative correction up to first order. We consider the cases of initial temperature $T_0=360$ MeV at initial proper time $\tau_0=0.6$ fm/c and $T_0=500$ MeV at $\tau_0=0.4$ fm/c, corresponding to the RHIC and the LHC initial conditions, respectively. The choice of two different initial conditions helps us to understand how the evolution of the system may vary under different initial conditions. 
We observe that the non-vanishing first order bulk viscous contribution to the entropy density is large, unlike in earlier studies where the first-order contribution vanishes as the mean field effects were absent~\cite{Jaiswal:2013fc}. This can be attributed to the fact that bulk correction to the distribution function explicitly depends on $z_k$ and constitutes term with $\delta\omega_k\Pi$ in the effective kinetic theory description of the medium. This dependency arises from the distribution function $f_k$, whose equilibrium part contains $z_k$ explicitly, and the evolution of the non-equilibrium part is affected due to the presence of the mean field force term in Boltzmann equation in Eq.~\eqref{26}. At later time, we see that as the system approaches equilibrium, the value of entropy density also approaches its equilibrium value. 
 
%%%%%%%%%%%%%%%%%%%%%%%%%%%%%%%%%%%%%%%%%%%%%%%%%%%%%%%%%%%%%%%%%%%%%
\section{Conclusion and Outlook}

In this paper, we have derived the first order dissipative hydrodynamic evolution equations within an effective covariant kinetic theory by realizing the system as a grand canonical ensemble of gluons and quarks, with a finite baryon chemical potential $\mu_q$ and non-zero quark mass $m_q$. The covariant effective kinetic theory is employed for the hot QCD matter within the EQPM. The thermal medium effects have been encoded through the EQPM by introducing the lattice equation of state in phase space momentum distribution through the effective fugacity parameter. We observed that the mean field contributions that emerge from the covariant kinetic theory induce sizable modification to the first order coefficients of the shear stress tensor, bulk viscous pressure and the particle diffusion of the hot QGP medium at non vanishing baryon chemical potential and non-zero quark mass in the temperature regime near to $T_c$. 
In the massless limit, our estimations at $\mu_q=0$ agree with the results of Ref.~\cite{Mitra:2018akk}, for the binary elastic collisions. 

We further studied the ratio of viscous coefficients $\frac{\beta_{\Pi}}{\beta_{\pi}}$ and compared the results with other parallel works. Furthermore, the relative significance of the charge conductivity and thermal conductivity with the viscous shear tensor have been investigated by evaluating the ratio $\frac{\beta_nT}{\beta_{\pi}}$ and $\frac{\kappa_q\mu^2_q}{\eta\pi^2T}$ respectively within RTA for different quark chemical potential. We found that at the lower temperature the charge conductivity is relatively smaller compared to the results at higher temperatures. Also, the effect of the baryon chemical potential is more visible in the temperature regime near to $T_c$. The proper time evolution of temperature and pressure anisotropy is seen to be sensitive to the viscous effects and the equation of state. Further, we investigated the time evolution of the Reynolds number and the viscous effects to the particle production measure in the medium. We studied the shear and bulk viscous corrections to the entropy four-current. We observe a non-vanishing first-order bulk viscous contribution to the entropy density within the effective kinetic theory description of medium. Finally, various predictions of the current work turned out to be consistent with the other parallel results.

The analysis presented in the manuscript is the first step towards the higher order (second and third) dissipative hydrodynamic evolution equation from the effective covariant kinetic theory within the EQPM. The investigation of the hydrodynamic evolution equations for the hot magnetized QGP medium (magnetohydrodynamics) would be another interesting problem to pursue. In addition, investigating the expanding medium with more realistic $3+1-$D hydrodynamics and the associated physical observables in heavy-ion collisions (such as heavy quark elliptic flow, dilepton spectra, etc.) is another direction worth investigating. We leave these problems for future work.

%%%%%%%%%%%%%%%%%%%%%%%%%%%%%%%%%%%%%%%%%%%%%%%%%%%%%%%%%%%%%%%%%%%%%

\begin{acknowledgments}
The authors would like to thank the anonymous referees of this article for comments which led to significant improvement in the manuscript.
S.~B. would like to acknowledge the hospitality of IIT Gandhinagar during the period of visit. M.~K. would like to acknowledge the hospitality of NISER Bhubaneswar. V.~C. would like to acknowledge SERB for the Early Career Research Award (ECRA/2016), and Department of Science and Technology (DST), Govt. of India for INSPIRE-Faculty Fellowship (IFA-13/PH-55). A.~J. is supported in part by the DST-INSPIRE faculty award under Grant No. DST/INSPIRE/04/2017/000038. We are indebted to the people of India for their generous support for research in basic sciences.
\end{acknowledgments}

%%%%%%%%%%%%%%%%%%%%%%%%%%%%%%%%%%%%%%%%%%%%%%%%%%%%%%%%%%%%%%%%%%%%%

\bigskip
\appendix

\section{Thermodynamic integrals }\label{A}

In this appendix section, we express the thermodynamic integrals used in the article explicitly in terms of other known functions. We have listed the expressions for both massive and massless cases.

In the derivation of viscous correction to the entropy current, the integrals of the form, $\tilde{J}^{(r)}_{k\, nm}$, are defined as:

\begin{align}\label{Jknmr def}
   \Tilde{J}^{(r)}_{k~nm}&=\frac{g_k}{2\pi^2}\frac{(-1)^m}{(2m+1)!!}\int_{0}^\infty{d\mid\vec{\Tilde{p}}_k\mid}~\big(u\cdot\Tilde{p}_k \big)^{n-2m-r-1}\nonumber\\
&\times\big(\mid\vec{\Tilde{p}}_k\mid\big)^{2m+2} f^{0}_k\, \tilde{f}^0_k .
\end{align}
%In addition to this, we also encountered integrals where the distribution function part of the integrand is $f^0_k\,\tilde{f}^0_k ({\tilde{f}}^0_k +a f^{0}_k)$ instead of $f^0_k\,\tilde{f}^0_k$ of \eqref{Jknmr def}. However those integrals can be expressed in terms of \eqref{Jknmr def} as well since we have, $\tilde{f}^0_k + a\,f^0_k = (1 - a\,f^0_k) + a\,f^0_k = 1$.

\subsection{Massive case}

For the case of massive quasipartons, the scalar thermodynamic integrals $\Tilde{J}^{(r)\pm}_{k~nm}$ and $\Tilde{L}^{(r)\pm}_{k~nm}$ can be expressed in terms of the modified Bessel function of second kind as shown in the following: 
\begin{widetext}

\begin{flalign}
    \Tilde{J}^{(1)+}_{q~42} =&\frac{g_qT^5y^5}{240\pi^2}\sum_{l=1}^{\infty}l(-1)^{l-1}z_q^l\cosh(l\alpha)\bigg[K_5(ly)-7K_3(ly)+22K_1(ly)+16K_{i,1}(ly)\bigg]\nonumber\\
    &-\delta\omega_q\frac{g_qT^4y^4}{60\pi^2}\sum_{l=1}^{\infty}l(-1)^{l-1}z_q^l\cosh(l\alpha)\bigg[K_4(ly)-8K_2(ly)+15K_0(ly)-8K_{i,2}(ly)\bigg],\label{A15}
\end{flalign}
\begin{flalign}
    \Tilde{J}^{(0)-}_{q~21}=&-\frac{g_qT^4y^4}{12\pi^2}\sum_{l=1}^{\infty}l(-1)^{l-1}z_q^l\sinh(l\alpha)\bigg[K_4(ly)-4K_2(ly)+3K_0(ly)\bigg]\nonumber\\
    &+\delta\omega_q\frac{g_qT^3y^3}{12\pi^2}\sum_{l=1}^{\infty}l(-1)^{l-1}z_q^l\sinh(l\alpha)\bigg[K_3(ly)-5K_1(ly)+4K_{i,1}(ly)\bigg],\label{A16}
\end{flalign}
\begin{flalign}
    \Tilde{J}^{(0)+}_{q~21}=&-\frac{g_q~T^4y^4}{12\pi^2}\sum_{l=1}^{\infty}l(-1)^{l-1}z_q^l\cosh(l\alpha)\bigg[K_4(ly)-4K_2(ly)+3K_0(ly)\bigg]\nonumber\\
    &+\delta\omega_q\frac{g_qT^3y^3}{12\pi^2}\sum_{l=1}^{\infty}l(-1)^{l-1}z_q^l\cosh(l\alpha)\bigg[K_3(ly)-5K_1(ly)+4K_{i,1}(ly)\bigg]\label{A17},
\end{flalign}
\begin{flalign}
    \Tilde{J}^{(1)+}_{q~21}=&-\frac{g_qT^3y^3}{12\pi^2}\sum_{l=1}^{\infty}l(-1)^{l-1}z_q^l\cosh(l\alpha)\bigg[K_3(ly)-5K_1(ly)+4K_{i,1}(ly)\bigg]\nonumber\\
    &+\delta\omega_q\frac{g_qT^2y^2}{3\pi^2}\sum_{l=1}^{\infty}l(-1)^{l-1}z_q^l\cosh(l\alpha)\bigg[K_2(ly)-3K_0(ly)+2K_{i,2}(ly)\bigg],\label{A18}
\end{flalign}
\begin{flalign}
    \Tilde{J}^{(0)+}_{q~31} =&-\frac{g_qT^5y^5}{48\pi^2}\sum_{l=1}^{\infty}l(-1)^{l-1}z_q^l\cosh(l\alpha)\bigg[K_5(ly)-3K_3(ly)+2K_{1}(ly)\bigg],\label{A19}
\end{flalign}    
\begin{flalign}
    \Tilde{J}^{(0)+}_{q~30} =&\frac{g_qT^5y^5}{16\pi^2}\sum_{l=1}^{\infty}l(-1)^{l-1}z_q^l\cosh(l\alpha)\bigg[K_5(ly)+ K_3(ly)- 2K_1(ly)\bigg]\nonumber\\
    &+\delta\omega_q\frac{g_qT^4y^4}{4\pi^2}\sum_{l=1}^{\infty}l(-1)^{l-1}z_q^l\cosh(l\alpha)\bigg[K_4(ly)- K_0(ly)\bigg],\label{A19.1}
\end{flalign}
\begin{flalign}
    \Tilde{J}^{(0)-}_{q~20}=&\frac{g_q T^4y^4}{8\pi^2}\sum_{l=1}^{\infty}l(-1)^{l-1}z_q^l\sinh(l\alpha)\bigg[K_4(ly)- K_0(ly)\bigg]+\delta\omega_q\frac{g_qT^3y^3}{4\pi^2}\sum_{l=1}^{\infty}l(-1)^{l-1}z_q^l\sinh(l\alpha)\bigg[K_3(ly)-K_1(ly)\bigg],\label{A19.2}
\end{flalign}
\begin{flalign}
    \Tilde{J}^{(0)+}_{q~10}=&\frac{g_q T^3y^3}{4\pi^2}\sum_{l=1}^{\infty}l(-1)^{l-1}z_q^l\sinh(l\alpha)\bigg[K_3(ly)- K_1(ly)\bigg],\label{A19.3}
\end{flalign}
\begin{flalign}
    \Tilde{J}^{(0)}_{q~30} =&\frac{g_q T^5y^5}{16\pi^2}\sum_{l=1}^{\infty}l(-1)^{l-1}z_g^l\cosh(l\alpha)\bigg[K_5(ly)+ K_3(ly)- 2K_1(ly)\bigg]\nonumber\\
    &+\delta\omega_q\frac{g_qT^4y^4}{4\pi^2}\sum_{l=1}^{\infty}l(-1)^{l-1}z_q^l\cosh(l\alpha)\bigg[K_4(ly)- K_0(ly)\bigg],\label{A19.4}
\end{flalign}
\begin{flalign}
    \Tilde{L}^{(1)+}_{q~42} =& \frac{g_qT^4y^4}{120\pi^2}\sum_{l=1}^{\infty}l(-1)^{l-1}z_q^l\cosh(l\alpha)\bigg[K_4(ly)-8K_2(ly)+15K_0(ly)-8K_{i,2}(ly)\bigg],\label{A20}
\end{flalign}
\begin{flalign}
    \Tilde{L}^{(0)+}_{q~31} =&-\frac{2\,g_q~T^4y^4}{24\pi^2}\sum_{l=1}^{\infty}l(-1)^{l-1}z_q^l\cosh(l\alpha)\bigg[K_4(ly)-4K_2(ly)+3K_0(ly)\bigg],\label{A21}
\end{flalign}
\begin{flalign}
    \Tilde{L}^{(0)-}_{q~21}=&-\frac{g_qT^3y^3}{12\pi^2}\sum_{l=1}^{\infty}l(-1)^{l-1}z_q^l\sinh(l\alpha)\bigg[K_3(ly)-5K_1(ly)+4K_{i,1}(ly)\bigg],\label{A22}
\end{flalign}    
\begin{flalign}
    \Tilde{L}^{(1)+}_{q~21}=&-\frac{g_qT^2y^2}{6\pi^2}\sum_{l=1}^{\infty}l(-1)^{l-1}z_q^l\cosh(l\alpha)\bigg[K_2(ly)-3K_0(ly)+2K_{i,2}(ly)\bigg], \label{A23}
\end{flalign}    

where the function $K_{i,n}(ly)$ is defined as, 
\begin{equation}
    K_{i,n}(ly)=\int_{0}^{\infty}{\frac{d\theta}{(\cosh{\theta})^n}\exp{(-ly\cosh{\theta})}}.\label{A24}
\end{equation}
\subsection{ Massless case }
  
For the massless case and non-vanishing baryon chemical potential, the thermodynamic integrals given in Eqs.~(\ref{34})-(\ref{36}) takes the following form,
%\begin{widetext}
\begin{align}
    \Tilde{J}^{(1)+}_{q~42} &=\frac{2\,g_q~T^5}{5\pi^2}\bigg[\!-2\bigg\{\mathrm{PolyLog}~[4,-e^{\alpha}z_q]+\mathrm{PolyLog}~[4,-e^{-\alpha}z_q]\bigg\}
    +\frac{\delta\omega_q}{T}\bigg\{\mathrm{PolyLog}~[3,-e^{-\alpha}z_q]+\mathrm{PolyLog}~[3,-e^{\alpha}z_q]\bigg\}\bigg],\label{A1}
\end{align}    
\begin{align}
    \Tilde{J}^{(0)-}_{q~21} &= \frac{g_qT^4}{\pi^2}\bigg[\bigg\{\mathrm{PolyLog}~[3,-e^{\alpha}z_q]-\mathrm{PolyLog}~[3,-e^{-\alpha}z_q]\bigg\}
    -\frac{\delta\omega_q}{3T}\bigg\{\mathrm{PolyLog}~[2,-e^{\alpha}z_q]-\mathrm{PolyLog}~[2,-e^{-\alpha}z_q]\bigg\}\bigg],\label{A2}
\end{align}    
\begin{align}   
    \Tilde{J}^{(0)+}_{q~21} &=\frac{g_qT^4}{\pi^2}\bigg[\bigg\{\mathrm{PolyLog}~[3,-e^{\alpha}z_q]+\mathrm{PolyLog}~[3,-e^{-\alpha}z_q]\bigg\}
    -\frac{\delta\omega_q}{3T}\bigg\{\mathrm{PolyLog}~[2,-e^{\alpha}z_q]+\mathrm{PolyLog}~[2,-e^{-\alpha}z_q]\bigg\}\bigg],\label{A3}
\end{align}    
\begin{align}    
    \Tilde{J}^{(1)+}_{q~21} &=\frac{g_qT^3}{3\pi^2}\bigg[\bigg\{\mathrm{PolyLog}~[2,-e^{-\alpha}z_q]+\mathrm{PolyLog}~[2,-e^{\alpha}z_q]\bigg\}
    +\frac{\delta\omega_q}{T}\bigg\{\mathrm{Log}~[1+e^{-\alpha}z_q]+\mathrm{Log}~[1+e^{\alpha}z_q]\bigg\}\bigg],\label{A4}
\end{align}    
\begin{align}    
    \Tilde{J}^{(0)+}_{q~31} &=-\frac{4\,g_q~T^5}{\pi^2}\bigg[\mathrm{PolyLog}~[4,-e^{-\alpha}z_q]+\mathrm{PolyLog}~[4,-e^{\alpha}z_q]\bigg],\label{A5}
\end{align}
\begin{align}
    \Tilde{J}^{(0)+}_{q~30} &=\frac{6 g_q T^5}{\pi^2}\bigg[\!2\bigg\{\mathrm{PolyLog}~[4,-e^{\alpha}z_q] - \mathrm{PolyLog}~[4,-e^{-\alpha}z_q] \bigg\}
    -\frac{\delta\omega_q}{T}\bigg\{\mathrm{PolyLog}~[3,-e^{-\alpha}z_q]+\mathrm{PolyLog}~[3,-e^{\alpha}z_q]\bigg\}\bigg],\label{A1.1}
\end{align}    
\begin{align}
    \Tilde{J}^{(0)-}_{q~20} &=\frac{g_q T^4}{\pi^2}\bigg[\!3\bigg\{\mathrm{PolyLog}~[3,-e^{-\alpha}z_q] - \mathrm{PolyLog}~[3,-e^{\alpha}z_q] \bigg\}
    + \frac{\delta\omega_q}{T}\bigg\{\mathrm{PolyLog}~[2,-e^{-\alpha}z_q]+\mathrm{PolyLog}~[2,-e^{\alpha}z_q]\bigg\}\bigg],\label{A1.2}
\end{align}
\begin{align}
    \Tilde{J}^{(0)+}_{q~10} &=-\frac{g_q T^3}{\pi^2}\bigg[\!\mathrm{PolyLog}~[2,-e^{-\alpha}z_q] + \mathrm{PolyLog}~[2,-e^{\alpha}z_q] \bigg],\label{A1.3}
\end{align}    
\begin{align}    
    \Tilde{L}^{(1)+}_{q~42} &= -\frac{g_qT^4}{5\pi^2}\bigg[\mathrm{PolyLog}~[3,-e^{-\alpha}z_q]+\mathrm{PolyLog}~[3,-e^{\alpha}z_q]\bigg],\label{A6}
\end{align}
\begin{align}
    \Tilde{L}^{(0)+}_{q~31} &=\frac{g_qT^4}{\pi^2}\bigg[\mathrm{PolyLog}~[3,-e^{-\alpha}z_q]+\mathrm{PolyLog}~[3,-e^{\alpha}z_q]\bigg],\label{A7}
\end{align}    
\begin{align}    
    \Tilde{L}^{(0)-}_{q~21} &=\frac{g_qT^3}{3\pi^2}\bigg[\mathrm{PolyLog}~[2,-e^{\alpha}z_q]-\mathrm{PolyLog}~[2,-e^{-\alpha}z_q]\bigg],\label{A8}
\end{align}    
\begin{align}    
    \Tilde{L}^{(1)+}_{q~21} &=-\frac{g_qT^2}{6\pi^2}\bigg[\mathrm{Log}~[1+e^{\alpha}z_q]+\mathrm{Log}~[1+e^{-\alpha}z_q]\bigg]\label{A9}
\end{align}
%\end{widetext}
For the gluonic case we obtain,
\begin{align}
    \Tilde{J}^{(1)}_{g~42} &=\frac{6\,g_gT^5}{15\pi^2}\bigg[2\,\mathrm{PolyLog}~[4,z_g]-\frac{\delta\omega_g}{T}\,\mathrm{PolyLog}~[3,z_g]\bigg],\label{A10}
    \end{align}
\begin{align}    
    \Tilde{J}^{(0)}_{g~31} &=-\frac{4\,g_g~T^5}{\pi^2}~\mathrm{PolyLog}~[4,z_g],\label{A11}
\end{align}  
\begin{align}  
    \Tilde{J}^{(0)}_{g~21} &=-\frac{g_gT^4}{\pi^2}~\mathrm{PolyLog}~[3,z_g],\label{A12}
\end{align}    
\begin{align}
    \Tilde{J}^{(0)}_{g~30} &=\frac{6 g_g T^5}{\pi^2}\bigg[\!~2~\mathrm{PolyLog}~[4,z_g] + \frac{\delta\omega_g}{T}\mathrm{PolyLog}~[3,z_g]\bigg],\label{A12.1}
\end{align}
\begin{align}
    \Tilde{L}^{(1)}_{g~42} &=\frac{g_gT^4}{5\pi^2}~\mathrm{PolyLog}~[3,z_g],\label{A13}
\end{align}
\begin{align}   
    \Tilde{L}^{(0)}_{g~31} &=-\frac{g_gT^4}{\pi^2}~\mathrm{PolyLog}~[3,z_g]\label{A14}.
\end{align} 
    
\end{widetext}

%%%%%%%%%%%%%%%%%%%%%%%%%%%%%%%%%%%%%%%%%%%%%%%%%%%%%%%%%%%%%%%%%%%%%

{}


\begin{thebibliography}{99}

%1
\bibitem{STAR}
  Adams et al. (STAR Collaboration), Nucl. Phys. A{\bf 757}, 102
  (2005); K. Adcox et al. (PHENIX Collaboration), Nucl.
  Phys. A{\bf 757}, 184 (2005); B.B. Back et al. (PHOBOS
  Collaboration), Nucl. Phys. A{\bf 757}, 28 (2005); A. Arsence
  et al. (BRAHMS Collaboration), Nucl. Phys. A{\bf 757}, 1 (2005).

%2\cite{Aamodt:2010pb}
\bibitem{Aamodt:2010pb} 
  K.~Aamodt {\it et al.} [ALICE Collaboration],
  %``Charged-particle multiplicity density at mid-rapidity in central Pb-Pb collisions at $\sqrt{s_{NN}} = 2.76$ TeV,''
  Phys.\ Rev.\ Lett.\  {\bf 105}, 252301 (2010).
    
 %\cite{Heinz:2008tv}
\bibitem{Heinz:2008tv} 
  U.~W.~Heinz,
  %``The Strongly coupled quark-gluon plasma created at RHIC,''
  J.\ Phys.\ A {\bf 42}, 214003 (2009)
 % doi:10.1088/1751-8113/42/21/214003
  [arXiv:0810.5529 [nucl-th]].
  %%CITATION = doi:10.1088/1751-8113/42/21/214003;%%
  %26 citations counted in INSPIRE as of 13 Feb 2019

%\cite{Jeon:2015dfa}
\bibitem{Jeon:2015dfa} 
  S.~Jeon and U.~Heinz,
  %``Introduction to Hydrodynamics,''
  Int.\ J.\ Mod.\ Phys.\ E {\bf 24}, no. 10, 1530010 (2015)
  %doi:10.1142/S0218301315300106
  [arXiv:1503.03931 [hep-ph]].
  %%CITATION = doi:10.1142/S0218301315300106;%%

%\cite{Florkowski:2017olj}
\bibitem{Florkowski:2017olj} 
  W.~Florkowski, M.~P.~Heller and M.~Spalinski,
  %``New theories of relativistic hydrodynamics in the LHC era,''
  Rept.\ Prog.\ Phys.\  {\bf 81}, no. 4, 046001 (2018)
  %doi:10.1088/1361-6633/aaa091
  [arXiv:1707.02282 [hep-ph]].
  %%CITATION = doi:10.1088/1361-6633/aaa091;%%

%\cite{Florkowski:2019qdp}
\bibitem{Florkowski:2019qdp} 
  W.~Florkowski, A.~Kumar, R.~Ryblewski and R.~Singh,
  %``Spin polarization evolution in a boost invariant hydrodynamical background,''
  Phys.\ Rev.\ C {\bf 99}, no. 4, 044910 (2019)
  %doi:10.1103/PhysRevC.99.044910
  [arXiv:1901.09655 [hep-ph]].
  %%CITATION = doi:10.1103/PhysRevC.99.044910;%%
  %3 citations counted in INSPIRE as of 21 Jun 2019

%\cite{Heinz:2013th}
\bibitem{Heinz:2013th} 
  U.~Heinz and R.~Snellings,
  %``Collective flow and viscosity in relativistic heavy-ion collisions,''
  Ann.\ Rev.\ Nucl.\ Part.\ Sci.\  {\bf 63}, 123 (2013)
  %doi:10.1146/annurev-nucl-102212-170540
  [arXiv:1301.2826 [nucl-th]].
  %%CITATION = doi:10.1146/annurev-nucl-102212-170540;%%

%\cite{Braun-Munzinger:2015hba}
\bibitem{Braun-Munzinger:2015hba} 
  P.~Braun-Munzinger, V.~Koch, T.~Schäfer and J.~Stachel,
  %``Properties of hot and dense matter from relativistic heavy ion collisions,''
  Phys.\ Rept.\  {\bf 621}, 76 (2016)
  %doi:10.1016/j.physrep.2015.12.003
  [arXiv:1510.00442 [nucl-th]].
  %%CITATION = doi:10.1016/j.physrep.2015.12.003;%%

%\cite{Jaiswal:2016hex}
\bibitem{Jaiswal:2016hex} 
  A.~Jaiswal and V.~Roy,
  %``Relativistic hydrodynamics in heavy-ion collisions: general aspects and recent developments,''
  Adv.\ High Energy Phys.\  {\bf 2016}, 9623034 (2016)
  %doi:10.1155/2016/9623034
  [arXiv:1605.08694 [nucl-th]].
  %%CITATION = doi:10.1155/2016/9623034;%%
  
 %\cite{Baier:2006um}
\bibitem{Baier:2006um} 
  R.~Baier, P.~Romatschke and U.~A.~Wiedemann,
  %``Dissipative hydrodynamics and heavy ion collisions,''
  Phys.\ Rev.\ C {\bf 73}, 064903 (2006)
%  doi:10.1103/PhysRevC.73.064903
  [hep-ph/0602249].
  %%CITATION = doi:10.1103/PhysRevC.73.064903;%%
  %223 citations counted in INSPIRE as of 13 Feb 2019 
  
  %\cite{Baier:2007ix}
\bibitem{Baier:2007ix} 
  R.~Baier, P.~Romatschke, D.~T.~Son, A.~O.~Starinets and M.~A.~Stephanov,
  %``Relativistic viscous hydrodynamics, conformal invariance, and holography,''
  JHEP {\bf 0804}, 100 (2008)
%  doi:10.1088/1126-6708/2008/04/100
  [arXiv:0712.2451 [hep-th]].
  %%CITATION = doi:10.1088/1126-6708/2008/04/100;%%
  %623 citations counted in INSPIRE as of 13 Feb 2019
    
%\cite{Denicol:2010xn}
\bibitem{Denicol:2010xn} 
  G.~S.~Denicol, T.~Koide and D.~H.~Rischke,
  %``Dissipative relativistic fluid dynamics: a new way to derive the equations of motion from kinetic theory,''
  Phys.\ Rev.\ Lett.\  {\bf 105}, 162501 (2010)
 % doi:10.1103/PhysRevLett.105.162501
  [arXiv:1004.5013 [nucl-th]].
  %%CITATION = doi:10.1103/PhysRevLett.105.162501;%%
  %183 citations counted in INSPIRE as of 13 Feb 2019
  
%\cite{Denicol:2012cn}
\bibitem{Denicol:2012cn} 
  G.~S.~Denicol, H.~Niemi, E.~Molnar and D.~H.~Rischke,
  %``Derivation of transient relativistic fluid dynamics from the Boltzmann equation,''
  Phys.\ Rev.\ D {\bf 85}, 114047 (2012)
%  Erratum: [Phys.\ Rev.\ D {\bf 91}, no. 3, 039902 (2015)]
%  doi:10.1103/PhysRevD.85.114047, 10.1103/PhysRevD.91.039902
  [arXiv:1202.4551 [nucl-th]].
  %%CITATION = doi:10.1103/PhysRevD.85.114047, 10.1103/PhysRevD.91.039902;%%
  %255 citations counted in INSPIRE as of 13 Feb 2019  
  
%\cite{Bhalerao:2013pza}
\bibitem{Bhalerao:2013pza} 
  R.~S.~Bhalerao, A.~Jaiswal, S.~Pal and V.~Sreekanth,
  %``Relativistic viscous hydrodynamics for heavy-ion collisions: A comparison between the Chapman-Enskog and Grad methods,''
  Phys.\ Rev.\ C {\bf 89}, no. 5, 054903 (2014)
 % doi:10.1103/PhysRevC.89.054903
  [arXiv:1312.1864 [nucl-th]].
  %%CITATION = doi:10.1103/PhysRevC.89.054903;%%
  %39 citations counted in INSPIRE as of 13 Feb 2019

%\cite{Luzum:2008cw}
\bibitem{Luzum:2008cw} 
  M.~Luzum and P.~Romatschke,
  %``Conformal Relativistic Viscous Hydrodynamics: Applications to RHIC results at s(NN)**(1/2) = 200-GeV,''
  Phys.\ Rev.\ C {\bf 78}, 034915 (2008)
%  Erratum: [Phys.\ Rev.\ C {\bf 79}, 039903 (2009)]
%  doi:10.1103/PhysRevC.78.034915, 10.1103/PhysRevC.79.039903
  [arXiv:0804.4015 [nucl-th]].
  %%CITATION = doi:10.1103/PhysRevC.78.034915, 10.1103/PhysRevC.79.039903;%%
  %737 citations counted in INSPIRE as of 13 Feb 2019  

  %\cite{ALICE:2016kpq}
\bibitem{ALICE:2016kpq} 
  J.~Adam {\it et al.} [ALICE Collaboration],
  %``Correlated event-by-event fluctuations of flow harmonics in Pb-Pb collisions at $\sqrt{s_{_{\rm NN}}}=2.76$ TeV,''
  Phys.\ Rev.\ Lett.\  {\bf 117}, 182301 (2016)
%  doi:10.1103/PhysRevLett.117.182301
  [arXiv:1604.07663 [nucl-ex]].
  %%CITATION = doi:10.1103/PhysRevLett.117.182301;%%
  %96 citations counted in INSPIRE as of 13 Feb 2019

%\cite{Adam:2016izf}
\bibitem{Adam:2016izf} 
  J.~Adam {\it et al.} [ALICE Collaboration],
  %``Anisotropic flow of charged particles in Pb-Pb collisions at $\sqrt{s_{\rm NN}}=5.02$ TeV,''
  Phys.\ Rev.\ Lett.\  {\bf 116}, no. 13, 132302 (2016)
%  doi:10.1103/PhysRevLett.116.132302
  [arXiv:1602.01119 [nucl-ex]].
  %%CITATION = doi:10.1103/PhysRevLett.116.132302;%%
  %100 citations counted in INSPIRE as of 13 Feb 2019

%\cite{Abelev:2013cva}
\bibitem{Abelev:2013cva} 
  B.~Abelev {\it et al.} [ALICE Collaboration],
  %``Directed Flow of Charged Particles at Midrapidity Relative to the Spectator Plane in Pb-Pb Collisions at $\sqrt{s_{NN}}$=2.76 TeV,''
  Phys.\ Rev.\ Lett.\  {\bf 111}, no. 23, 232302 (2013)
%  doi:10.1103/PhysRevLett.111.232302
  [arXiv:1306.4145 [nucl-ex]].
  %%CITATION = doi:10.1103/PhysRevLett.111.232302;%%
  %53 citations counted in INSPIRE as of 13 Feb 2019  
  
%\cite{Adam:2016nfo}
\bibitem{Adam:2016nfo} 
  J.~Adam {\it et al.} [ALICE Collaboration],
  %``Higher harmonic flow coefficients of identified hadrons in Pb-Pb collisions at $\sqrt{s_{\rm NN}}$ = 2.76 TeV,''
  JHEP {\bf 1609}, 164 (2016)
%  doi:10.1007/JHEP09(2016)164
  [arXiv:1606.06057 [nucl-ex]].
  %%CITATION = doi:10.1007/JHEP09(2016)164;%%
  %34 citations counted in INSPIRE as of 13 Feb 2019  
  
%\cite{Arnold:2000dr}
\bibitem{Arnold:2000dr} 
  P.~B.~Arnold, G.~D.~Moore and L.~G.~Yaffe,
  %``Transport coefficients in high temperature gauge theories. 1. Leading log results,''
  JHEP {\bf 0011}, 001 (2000)
  %doi:10.1088/1126-6708/2000/11/001
  [hep-ph/0010177].
  %%CITATION = doi:10.1088/1126-6708/2000/11/001;%%
  %574 citations counted in INSPIRE as of 12 Jul 2019  
  
%\cite{Arnold:2003zc}
\bibitem{Arnold:2003zc} 
  P.~B.~Arnold, G.~D.~Moore and L.~G.~Yaffe,
  %``Transport coefficients in high temperature gauge theories. 2. Beyond leading log,''
  JHEP {\bf 0305}, 051 (2003)
  %doi:10.1088/1126-6708/2003/05/051
  [hep-ph/0302165].
  %%CITATION = doi:10.1088/1126-6708/2003/05/051;%%
  %537 citations counted in INSPIRE as of 12 Jul 2019
  
  %\cite{Meyer:2007ic}
\bibitem{Meyer:2007ic} 
  H.~B.~Meyer,
  %``A Calculation of the shear viscosity in SU(3) gluodynamics,''
  Phys.\ Rev.\ D {\bf 76}, 101701 (2007)
 % doi:10.1103/PhysRevD.76.101701
  [arXiv:0704.1801 [hep-lat]].
  %%CITATION = doi:10.1103/PhysRevD.76.101701;%%
  %350 citations counted in INSPIRE as of 13 Feb 2019
  
%\cite{Meyer:2007dy}
\bibitem{Meyer:2007dy} 
  H.~B.~Meyer,
  %``A Calculation of the bulk viscosity in SU(3) gluodynamics,''
  Phys.\ Rev.\ Lett.\  {\bf 100}, 162001 (2008)
 % doi:10.1103/PhysRevLett.100.162001
  [arXiv:0710.3717 [hep-lat]].
  %%CITATION = doi:10.1103/PhysRevLett.100.162001;%%
  %254 citations counted in INSPIRE as of 13 Feb 2019  
 
%\cite{Bluhm:2011xu}
\bibitem{Bluhm:2011xu} 
  M.~Bluhm, B.~Kampfer and K.~Redlich,
  %``Ratio of bulk to shear viscosity in a quasigluon plasma: from weak to strong coupling,''
  Phys.\ Lett.\ B {\bf 709}, 77 (2012),
  [arXiv:1101.3072 [hep-ph]].
  %%CITATION = doi:10.1016/j.physletb.2012.01.069;%%
  %17 citations counted in INSPIRE as of 23 Nov 2018 
  
%\cite{Deb:2016myz}
\bibitem{Deb:2016myz} 
  P.~Deb, G.~P.~Kadam and H.~Mishra,
  %``Estimating transport coefficients in hot and dense quark matter,''
  Phys.\ Rev.\ D {\bf 94}, no. 9, 094002 (2016)
 % doi:10.1103/PhysRevD.94.094002
  [arXiv:1603.01952 [hep-ph]].
  %%CITATION = doi:10.1103/PhysRevD.94.094002;%%
  %18 citations counted in INSPIRE as of 13 Feb 2019  
 
 %\cite{Ghosh:2015mda}
\bibitem{Ghosh:2015mda} 
  S.~Ghosh, T.~C.~Peixoto, V.~Roy, F.~E.~Serna and G.~Krein,
  %``Shear and bulk viscosities of quark matter from quark-meson fluctuations in the Nambu–Jona-Lasinio model,''
  Phys.\ Rev.\ C {\bf 93}, no. 4, 045205 (2016)
%  doi:10.1103/PhysRevC.93.045205
  [arXiv:1507.08798 [nucl-th]].
  %%CITATION = doi:10.1103/PhysRevC.93.045205;%%
  %24 citations counted in INSPIRE as of 13 Feb 2019 
  
 %\cite{Mitra:2016zdw}
\bibitem{Mitra:2016zdw} 
  S.~Mitra and V.~Chandra,
  %``Thermal relaxation, electrical conductivity, and charge diffusion in a hot QCD medium,''
  Phys.\ Rev.\ D {\bf 94}, no. 3, 034025 (2016)
%  doi:10.1103/PhysRevD.94.034025
  [arXiv:1606.08556 [nucl-th]].
  %%CITATION = doi:10.1103/PhysRevD.94.034025;%%
  %20 citations counted in INSPIRE as of 13 Feb 2019
  
%\cite{Jaiswal:2014isa}
\bibitem{Jaiswal:2014isa} 
  A.~Jaiswal, R.~Ryblewski and M.~Strickland,
  %``Transport coefficients for bulk viscous evolution in the relaxation time approximation,''
  Phys.\ Rev.\ C {\bf 90}, 044908 (2014)
  %doi:10.1103/PhysRevC.90.044908
  [arXiv:1407.7231 [hep-ph]].
  %%CITATION = doi:10.1103/PhysRevC.90.044908;%%  
  
%\cite{Florkowski:2015lra}
\bibitem{Florkowski:2015lra} 
  W.~Florkowski, A.~Jaiswal, E.~Maksymiuk, R.~Ryblewski and M.~Strickland,
  %``Relativistic quantum transport coefficients for second-order viscous hydrodynamics,''
  Phys.\ Rev.\ C {\bf 91}, 054907 (2015)
  %doi:10.1103/PhysRevC.91.054907
  [arXiv:1503.03226 [nucl-th]].
  %%CITATION = doi:10.1103/PhysRevC.91.054907;%%
  
%\cite{Marty:2013ita}
\bibitem{Marty:2013ita} 
  R.~Marty, E.~Bratkovskaya, W.~Cassing, J.~Aichelin and H.~Berrehrah,
  %``Transport coefficients from the Nambu-Jona-Lasinio model for $SU(3)_f$,''
  Phys.\ Rev.\ C {\bf 88}, 045204 (2013)
%  doi:10.1103/PhysRevC.88.045204
  [arXiv:1305.7180 [hep-ph]].
  %%CITATION = doi:10.1103/PhysRevC.88.045204;%%
  %83 citations counted in INSPIRE as of 13 Feb 2019  
  
%\cite{Mitra:2017sjo}
\bibitem{Mitra:2017sjo} 
  S.~Mitra and V.~Chandra,
  %``Transport coefficients of a hot QCD medium and their relative significance in heavy-ion collisions,''
  Phys.\ Rev.\ D {\bf 96}, no. 9, 094003 (2017)
%  doi:10.1103/PhysRevD.96.094003
  [arXiv:1702.05728 [nucl-th]].
  %%CITATION = doi:10.1103/PhysRevD.96.094003;%%
  %11 citations counted in INSPIRE as of 13 Feb 2019  
  
%\cite{Niemi:2011ix}
\bibitem{Niemi:2011ix} 
  H.~Niemi, G.~S.~Denicol, P.~Huovinen, E.~Molnar and D.~H.~Rischke,
  %``Influence of the shear viscosity of the quark-gluon plasma on elliptic flow in ultrarelativistic heavy-ion collisions,''
  Phys.\ Rev.\ Lett.\  {\bf 106}, 212302 (2011)
%  doi:10.1103/PhysRevLett.106.212302
  [arXiv:1101.2442 [nucl-th]].
  %%CITATION = doi:10.1103/PhysRevLett.106.212302;%%
  %205 citations counted in INSPIRE as of 13 Feb 2019 
 
%\cite{Niemi:2012ry}
\bibitem{Niemi:2012ry} 
  H.~Niemi, G.~S.~Denicol, P.~Huovinen, E.~Molnar and D.~H.~Rischke,
  %``Influence of a temperature-dependent shear viscosity on the azimuthal asymmetries of transverse momentum spectra in ultrarelativistic heavy-ion collisions,''
  Phys.\ Rev.\ C {\bf 86}, 014909 (2012)
%  doi:10.1103/PhysRevC.86.014909
  [arXiv:1203.2452 [nucl-th]].
  %%CITATION = doi:10.1103/PhysRevC.86.014909;%%
  %115 citations counted in INSPIRE as of 13 Feb 2019 
  
 %\cite{Romatschke:2007mq}
\bibitem{Romatschke:2007mq} 
  P.~Romatschke and U.~Romatschke,
  %``Viscosity Information from Relativistic Nuclear Collisions: How Perfect is the Fluid Observed at RHIC?,''
  Phys.\ Rev.\ Lett.\  {\bf 99}, 172301 (2007)
 % doi:10.1103/PhysRevLett.99.172301
  [arXiv:0706.1522 [nucl-th]].
  %%CITATION = doi:10.1103/PhysRevLett.99.172301;%%
  %839 citations counted in INSPIRE as of 13 Feb 2019
  
%\cite{Song:2008hj}
\bibitem{Song:2008hj} 
  H.~Song and U.~W.~Heinz,
  %``Extracting the QGP viscosity from RHIC data - A Status report from viscous hydrodynamics,''
  J.\ Phys.\ G {\bf 36}, 064033 (2009)
 % doi:10.1088/0954-3899/36/6/064033
  [arXiv:0812.4274 [nucl-th]].
  %%CITATION = doi:10.1088/0954-3899/36/6/064033;%%
  %188 citations counted in INSPIRE as of 13 Feb 2019
  
%\cite{Song:2010aq}
\bibitem{Song:2010aq} 
  H.~Song, S.~A.~Bass and U.~Heinz,
  %``Viscous QCD matter in a hybrid hydrodynamic+Boltzmann approach,''
  Phys.\ Rev.\ C {\bf 83}, 024912 (2011)
 % doi:10.1103/PhysRevC.83.024912
  [arXiv:1012.0555 [nucl-th]].
  %%CITATION = doi:10.1103/PhysRevC.83.024912;%%
  %135 citations counted in INSPIRE as of 13 Feb 2019
 
 %\cite{Song:2011qa}
\bibitem{Song:2011qa} 
  H.~Song, S.~A.~Bass and U.~Heinz,
  %``Elliptic flow in 200 A GeV Au+Au collisions and 2.76 A TeV Pb+Pb collisions: insights from viscous hydrodynamics + hadron cascade hybrid model,''
  Phys.\ Rev.\ C {\bf 83}, 054912 (2011)
%  Erratum: [Phys.\ Rev.\ C {\bf 87}, no. 1, 019902 (2013)]
%  doi:10.1103/PhysRevC.83.054912, 10.1103/PhysRevC.87.019902
  [arXiv:1103.2380 [nucl-th]].
  %%CITATION = doi:10.1103/PhysRevC.83.054912, 10.1103/PhysRevC.87.019902;%%
  %140 citations counted in INSPIRE as of 13 Feb 2019 
 
 %\cite{Song:2011hk}
\bibitem{Song:2011hk} 
  H.~Song, S.~A.~Bass, U.~Heinz, T.~Hirano and C.~Shen,
  %``Hadron spectra and elliptic flow for 200 A GeV Au+Au collisions from viscous hydrodynamics coupled to a Boltzmann cascade,''
  Phys.\ Rev.\ C {\bf 83}, 054910 (2011)
 % Erratum: [Phys.\ Rev.\ C {\bf 86}, 059903 (2012)]
 % doi:10.1103/PhysRevC.83.054910, 10.1103/PhysRevC.86.059903
  [arXiv:1101.4638 [nucl-th]].
  %%CITATION = doi:10.1103/PhysRevC.83.054910, 10.1103/PhysRevC.86.059903;%%
  %132 citations counted in INSPIRE as of 13 Feb 2019 
  
%\cite{Schenke:2010rr}
\bibitem{Schenke:2010rr} 
  B.~Schenke, S.~Jeon and C.~Gale,
  %``Elliptic and triangular flow in event-by-event (3+1)D viscous hydrodynamics,''
  Phys.\ Rev.\ Lett.\  {\bf 106}, 042301 (2011)
%  doi:10.1103/PhysRevLett.106.042301
  [arXiv:1009.3244 [hep-ph]].
  %%CITATION = doi:10.1103/PhysRevLett.106.042301;%%
  %488 citations counted in INSPIRE as of 13 Feb 2019 
  
 %\cite{Gale:2012rq}
\bibitem{Gale:2012rq} 
  C.~Gale, S.~Jeon, B.~Schenke, P.~Tribedy and R.~Venugopalan,
  %``Event-by-event anisotropic flow in heavy-ion collisions from combined Yang-Mills and viscous fluid dynamics,''
  Phys.\ Rev.\ Lett.\  {\bf 110}, no. 1, 012302 (2013)
%  doi:10.1103/PhysRevLett.110.012302
  [arXiv:1209.6330 [nucl-th]].
  %%CITATION = doi:10.1103/PhysRevLett.110.012302;%%
  %402 citations counted in INSPIRE as of 13 Feb 2019 
  
%\cite{Ryu:2015vwa}
\bibitem{Ryu:2015vwa} 
  S.~Ryu, J.-F.~Paquet, C.~Shen, G.~S.~Denicol, B.~Schenke, S.~Jeon and C.~Gale,
  %``Importance of the Bulk Viscosity of QCD in Ultrarelativistic Heavy-Ion Collisions,''
  Phys.\ Rev.\ Lett.\  {\bf 115}, no. 13, 132301 (2015)
%  doi:10.1103/PhysRevLett.115.132301
  [arXiv:1502.01675 [nucl-th]].
  %%CITATION = doi:10.1103/PhysRevLett.115.132301;%%
  %164 citations counted in INSPIRE as of 13 Feb 2019  

%\cite{Huang:2010sa}
\bibitem{Huang:2010sa} 
  X.~G.~Huang, T.~Kodama, T.~Koide and D.~H.~Rischke,
  %``Bulk Viscosity and Relaxation Time of Causal Dissipative Relativistic Fluid Dynamics,''
  Phys.\ Rev.\ C {\bf 83}, 024906 (2011)
%  doi:10.1103/PhysRevC.83.024906
  [arXiv:1010.4359 [nucl-th]].
  %%CITATION = doi:10.1103/PhysRevC.83.024906;%%
  %24 citations counted in INSPIRE as of 13 Feb 2019  
  
%\cite{Denicol:2014vaa}
\bibitem{Denicol:2014vaa} 
  G.~S.~Denicol, S.~Jeon and C.~Gale,
  %``Transport Coefficients of Bulk Viscous Pressure in the 14-moment approximation,''
  Phys.\ Rev.\ C {\bf 90}, no. 2, 024912 (2014)
%  doi:10.1103/PhysRevC.90.024912
  [arXiv:1403.0962 [nucl-th]].
  %%CITATION = doi:10.1103/PhysRevC.90.024912;%%
  %80 citations counted in INSPIRE as of 13 Feb 2019  

%\cite{Dobado:2011qu}
\bibitem{Dobado:2011qu} 
  A.~Dobado, F.~J.~Llanes-Estrada and J.~M.~Torres-Rincon,
  %``Bulk viscosity of low-temperature strongly interacting matter,''
  Phys.\ Lett.\ B {\bf 702}, 43 (2011)
%  doi:10.1016/j.physletb.2011.06.059
  [arXiv:1103.0735 [hep-ph]].
  %%CITATION = doi:10.1016/j.physletb.2011.06.059;%%
  %33 citations counted in INSPIRE as of 13 Feb 2019

%\cite{Bluhm:2010qf}
\bibitem{Bluhm:2010qf} 
  M.~Bluhm, B.~Kampfer and K.~Redlich,
  %``Bulk and shear viscosities of the gluon plasma in a quasiparticle description,''
  Phys.\ Rev.\ C {\bf 84}, 025201 (2011)
%  doi:10.1103/PhysRevC.84.025201
  [arXiv:1011.5634 [hep-ph]].
  %%CITATION = doi:10.1103/PhysRevC.84.025201;%%
  %64 citations counted in INSPIRE as of 13 Feb 2019
  
%\cite{Romatschke:2011qp}
\bibitem{Romatschke:2011qp} 
  P.~Romatschke,
  %``Relativistic (Lattice) Boltzmann Equation with Non-Ideal Equation of State,''
  Phys.\ Rev.\ D {\bf 85}, 065012 (2012)
  %doi:10.1103/PhysRevD.85.065012
  [arXiv:1108.5561 [gr-qc]].
  %%CITATION = doi:10.1103/PhysRevD.85.065012;%%
  %59 citations counted in INSPIRE as of 12 Jul 2019  
  
%\cite{Tinti:2016bav}
\bibitem{Tinti:2016bav} 
  L.~Tinti, A.~Jaiswal and R.~Ryblewski,
  %``Quasiparticle second-order viscous hydrodynamics from kinetic theory,''
  Phys.\ Rev.\ D {\bf 95}, no. 5, 054007 (2017)
%  doi:10.1103/PhysRevD.95.054007
  [arXiv:1612.07329 [nucl-th]].
  %%CITATION = doi:10.1103/PhysRevD.95.054007;%%
  %18 citations counted in INSPIRE as of 13 Feb 2019  
  
%\cite{Alqahtani:2016rth}
\bibitem{Alqahtani:2016rth} 
  M.~Alqahtani, M.~Nopoush and M.~Strickland,
  %``Quasiparticle anisotropic hydrodynamics for central collisions,''
  Phys.\ Rev.\ C {\bf 95}, no. 3, 034906 (2017)
 % doi:10.1103/PhysRevC.95.034906
  [arXiv:1605.02101 [nucl-th]].
  %%CITATION = doi:10.1103/PhysRevC.95.034906;%%
  %24 citations counted in INSPIRE as of 13 Feb 2019  
  
 %\cite{Alqahtani:2017jwl}
\bibitem{Alqahtani:2017jwl} 
  M.~Alqahtani, M.~Nopoush, R.~Ryblewski and M.~Strickland,
  %``(3+1)D Quasiparticle Anisotropic Hydrodynamics for Ultrarelativistic Heavy-Ion Collisions,''
  Phys.\ Rev.\ Lett.\  {\bf 119}, no. 4, 042301 (2017)
 % doi:10.1103/PhysRevLett.119.042301
  [arXiv:1703.05808 [nucl-th]].
  %%CITATION = doi:10.1103/PhysRevLett.119.042301;%%
  %33 citations counted in INSPIRE as of 13 Feb 2019 
  
%\cite{Kurian:2018dbn}
\bibitem{Kurian:2018dbn} 
  M.~Kurian and V.~Chandra,
  %``Bulk viscosity of a hot QCD medium in a strong magnetic field within the relaxation-time approximation,''
  Phys.\ Rev.\ D {\bf 97}, no. 11, 116008 (2018)
%  doi:10.1103/PhysRevD.97.116008
  [arXiv:1802.07904 [nucl-th]].
  %%CITATION = doi:10.1103/PhysRevD.97.116008;%%
  %5 citations counted in INSPIRE as of 13 Feb 2019

%\cite{Kurian:2017yxj}
\bibitem{Kurian:2017yxj} 
  M.~Kurian and V.~Chandra,
  %``Effective description of hot QCD medium in strong magnetic field and longitudinal conductivity,''
  Phys.\ Rev.\ D {\bf 96}, no. 11, 114026 (2017)
 % doi:10.1103/PhysRevD.96.114026
  [arXiv:1709.08320 [nucl-th]].
  %%CITATION = doi:10.1103/PhysRevD.96.114026;%%
  %12 citations counted in INSPIRE as of 13 Feb 2019  
  
%\cite{Kurian:2018qwb}
\bibitem{Kurian:2018qwb} 
  M.~Kurian, S.~Mitra, S.~Ghosh and V.~Chandra,
  %``Transport coefficients of hot magnetized QCD matter beyond the lowest Landau level approximation,''
  arXiv:1805.07313 [nucl-th].
  %%CITATION = ARXIV:1805.07313;%%
  %6 citations counted in INSPIRE as of 13 Feb 2019
  
 %\cite{Rozynek:2018tev}
\bibitem{Rozynek:2018tev} 
  J.~Rożynek and G.~Wilk,
  %``Nonextensive quasiparticle description of QCD matter,''
  arXiv:1810.07008 [hep-ph].
  %%CITATION = ARXIV:1810.07008;%%
  
 %\cite{Chandra:2011en}
\bibitem{Chandra:2011en} 
  V.~Chandra and V.~Ravishankar,
  %``A quasi-particle description of $(2+1)$- flavor lattice QCD equation of state,''
  Phys.\ Rev.\ D {\bf 84}, 074013 (2011)
 % doi:10.1103/PhysRevD.84.074013
  [arXiv:1103.0091 [nucl-th]].
  %%CITATION = doi:10.1103/PhysRevD.84.074013;%%
  %32 citations counted in INSPIRE as of 13 Feb 2019 
  
%\cite{Chandra:2007ca}
\bibitem{Chandra:2007ca} 
  V.~Chandra, R.~Kumar and V.~Ravishankar,
  %``Hot QCD equation of state and relativistic heavy ion collisions,''
  Phys.\ Rev.\ C {\bf 76}, 054909 (2007)
 % Erratum: [Phys.\ Rev.\ C {\bf 76}, 069904 (2007)]
%  doi:10.1103/PhysRevC.76.069904, 10.1103/PhysRevC.76.054909
  [arXiv:0705.2690 [nucl-th]].
  %%CITATION = doi:10.1103/PhysRevC.76.069904, 10.1103/PhysRevC.76.054909;%%
  %32 citations counted in INSPIRE as of 13 Feb 2019  
  
 %\cite{Mitra:2018akk}
\bibitem{Mitra:2018akk} 
  S.~Mitra and V.~Chandra,
  %``Covariant kinetic theory for effective fugacity quasiparticle model and first order transport coefficients for hot QCD matter,''
  Phys.\ Rev.\ D {\bf 97}, no. 3, 034032 (2018)
 % doi:10.1103/PhysRevD.97.034032
  [arXiv:1801.01700 [nucl-th]].
  %%CITATION = doi:10.1103/PhysRevD.97.034032;%%
  %4 citations counted in INSPIRE as of 13 Feb 2019 

%\cite{El:2009vj}
\bibitem{El:2009vj} 
  A.~El, Z.~Xu and C.~Greiner,
  %``Third-order relativistic dissipative hydrodynamics,''
  Phys.\ Rev.\ C {\bf 81}, 041901 (2010)
  %doi:10.1103/PhysRevC.81.041901
  [arXiv:0907.4500 [hep-ph]].
  %%CITATION = doi:10.1103/PhysRevC.81.041901;%%
  %103 citations counted in INSPIRE as of 07 Dec 2019
 
 %\cite{Chattopadhyay:2014lya}
\bibitem{Chattopadhyay:2014lya} 
  C.~Chattopadhyay, A.~Jaiswal, S.~Pal and R.~Ryblewski,
  %``Relativistic third-order viscous corrections to the entropy four-current from kinetic theory,''
  Phys.\ Rev.\ C {\bf 91}, no. 2, 024917 (2015)
  %doi:10.1103/PhysRevC.91.024917
  [arXiv:1411.2363 [nucl-th]].
  %%CITATION = doi:10.1103/PhysRevC.91.024917;%%
  %29 citations counted in INSPIRE as of 07 Dec 2019
  
 \bibitem{kamf}
V. Goloviznin and H. Satz, Z. Phys. C {\bf 57}, 671 (1994).

%\cite{Peshier:1995ty}
\bibitem{Peshier:1995ty} 
  A.~Peshier, B.~Kampfer, O.~P.~Pavlenko and G.~Soff,
  %``A Massive quasiparticle model of the SU(3) gluon plasma,''
  Phys.\ Rev.\ D {\bf 54}, 2399 (1996).
 % doi:10.1103/PhysRevD.54.2399
  %%CITATION = doi:10.1103/PhysRevD.54.2399;%%
  %281 citations counted in INSPIRE as of 14 Feb 2019
  
  %\cite{DElia:1997sdk}
\bibitem{DElia:1997sdk} 
  M.~D'Elia, A.~Di Giacomo and E.~Meggiolaro,
  %``Field strength correlators in full QCD,''
  Phys.\ Lett.\ B {\bf 408}, 315 (1997)
%  doi:10.1016/S0370-2693(97)00814-9
  [hep-lat/9705032].
  %%CITATION = doi:10.1016/S0370-2693(97)00814-9;%%
  %249 citations counted in INSPIRE as of 14 Feb 2019
  
%\cite{DElia:2002hkf}
\bibitem{DElia:2002hkf} 
  M.~D'Elia, A.~Di Giacomo and E.~Meggiolaro,
  %``Gauge invariant field strength correlators in pure Yang-Mills and full QCD at finite temperature,''
  Phys.\ Rev.\ D {\bf 67}, 114504 (2003)
%  doi:10.1103/PhysRevD.67.114504
  [hep-lat/0205018].
  %%CITATION = doi:10.1103/PhysRevD.67.114504;%%
  %128 citations counted in INSPIRE as of 14 Feb 2019
  
  %\cite{Castorina:2007qv}
\bibitem{Castorina:2007qv} 
  P.~Castorina and M.~Mannarelli,
  %``Effective degrees of freedom and gluon condensation in the high temperature deconfined phase,''
  Phys.\ Rev.\ C {\bf 75}, 054901 (2007)
%  doi:10.1103/PhysRevC.75.054901
  [hep-ph/0701206 [HEP-PH]].
  %%CITATION = doi:10.1103/PhysRevC.75.054901;%%
  %41 citations counted in INSPIRE as of 14 Feb 2019

%\cite{Castorina:2005wi}
\bibitem{Castorina:2005wi} 
  P.~Castorina and M.~Mannarelli,
  %``Effective degrees of freedom of the quark-gluon plasma,''
  Phys.\ Lett.\ B {\bf 644}, 336 (2007)
%  doi:10.1016/j.physletb.2006.11.058
  [hep-ph/0510349].
  %%CITATION = doi:10.1016/j.physletb.2006.11.058;%%
  %21 citations counted in INSPIRE as of 14 Feb 2019  
  
 %\cite{Bannur:2006js}
\bibitem{Bannur:2006js} 
  V.~M.~Bannur,
  %``Self-consistent quasiparticle model for quark-gluon plasma,''
  Phys.\ Rev.\ C {\bf 75}, 044905 (2007)
 % doi:10.1103/PhysRevC.75.044905
  [hep-ph/0609188].
  %%CITATION = doi:10.1103/PhysRevC.75.044905;%%
  %40 citations counted in INSPIRE as of 13 Feb 2019
 
%24\cite{Koothottil:2018akg}
\bibitem{Koothottil:2018akg} 
  S.~Koothottil and V.~M.~Bannur,
  %``Thermodynamic Behaviour of Magnetized QGP within the Self-Consistent Quasiparticle Model,''
 [arXiv:1811.05377 [nucl-th]].
  %%CITATION = ARXIV:1811.05377;%%
 
 %\cite{Dumitru:2001xa}
\bibitem{Dumitru:2001xa} 
  A.~Dumitru and R.~D.~Pisarski,
  %``Degrees of freedom and the deconfining phase transition,''
  Phys.\ Lett.\ B {\bf 525}, 95 (2002)
%  doi:10.1016/S0370-2693(01)01424-1
  [hep-ph/0106176].
  %%CITATION = doi:10.1016/S0370-2693(01)01424-1;%%
  %115 citations counted in INSPIRE as of 13 Feb 2019 
  
%\cite{Fukushima:2003fw}
\bibitem{Fukushima:2003fw} 
  K.~Fukushima,
  %``Chiral effective model with the Polyakov loop,''
  Phys.\ Lett.\ B {\bf 591}, 277 (2004)
%  doi:10.1016/j.physletb.2004.04.027
  [hep-ph/0310121].
  %%CITATION = doi:10.1016/j.physletb.2004.04.027;%%
  %745 citations counted in INSPIRE as of 13 Feb 2019
  
%\cite{Ghosh:2006qh}
\bibitem{Ghosh:2006qh} 
  S.~K.~Ghosh, T.~K.~Mukherjee, M.~G.~Mustafa and R.~Ray,
  %``Susceptibilities and speed of sound from PNJL model,''
  Phys.\ Rev.\ D {\bf 73}, 114007 (2006)
 % doi:10.1103/PhysRevD.73.114007
  [hep-ph/0603050].
  %%CITATION = doi:10.1103/PhysRevD.73.114007;%%
  %198 citations counted in INSPIRE as of 13 Feb 2019  
 
%\cite{Su:2014rma}
\bibitem{Su:2014rma} 
  N.~Su and K.~Tywoniuk,
  %``Massless Mode and Positivity Violation in Hot QCD,''
  Phys.\ Rev.\ Lett.\  {\bf 114}, no. 16, 161601 (2015)
%  doi:10.1103/PhysRevLett.114.161601
  [arXiv:1409.3203 [hep-ph]].
  %%CITATION = doi:10.1103/PhysRevLett.114.161601;%%
  %40 citations counted in INSPIRE as of 13 Feb 2019
  
%\cite{Florkowski:2015dmm}
\bibitem{Florkowski:2015dmm} 
  W.~Florkowski, R.~Ryblewski, N.~Su and K.~Tywoniuk,
  %``Transport coefficients of the Gribov-Zwanziger plasma,''
  Phys.\ Rev.\ C {\bf 94}, no. 4, 044904 (2016)
 % doi:10.1103/PhysRevC.94.044904
  [arXiv:1509.01242 [hep-ph]].
  %%CITATION = doi:10.1103/PhysRevC.94.044904;%%
  %25 citations counted in INSPIRE as of 13 Feb 2019

%\cite{Bandyopadhyay:2015wua}
\bibitem{Bandyopadhyay:2015wua} 
  A.~Bandyopadhyay, N.~Haque, M.~G.~Mustafa and M.~Strickland,
  %``Dilepton rate and quark number susceptibility with the Gribov action,''
  Phys.\ Rev.\ D {\bf 93}, no. 6, 065004 (2016)
  %doi:10.1103/PhysRevD.93.065004
  [arXiv:1508.06249 [hep-ph]].
  %%CITATION = doi:10.1103/PhysRevD.93.065004;%%
  %21 citations counted in INSPIRE as of 13 Feb 2019
  
  
%\cite{Cheng:2007jq}
\bibitem{Cheng:2007jq} 
  M.~Cheng {\it et al.},
  %``The QCD equation of state with almost physical quark masses,''
  Phys.\ Rev.\ D {\bf 77}, 014511 (2008)
 % doi:10.1103/PhysRevD.77.014511
  [arXiv:0710.0354 [hep-lat]].
  %%CITATION = doi:10.1103/PhysRevD.77.014511;%%
  %573 citations counted in INSPIRE as of 13 Feb 2019

%\cite{Borsanyi:2013bia}
\bibitem{Borsanyi:2013bia} 
  S.~Borsanyi, Z.~Fodor, C.~Hoelbling, S.~D.~Katz, S.~Krieg and K.~K.~Szabo,
  %``Full result for the QCD equation of state with 2+1 flavors,''
  Phys.\ Lett.\ B {\bf 730}, 99 (2014)
 % doi:10.1016/j.physletb.2014.01.007
  [arXiv:1309.5258 [hep-lat]].
  %%CITATION = doi:10.1016/j.physletb.2014.01.007;%%
  %462 citations counted in INSPIRE as of 13 Feb 2019
  
\bibitem{vogt}   
R. Vogt,{\it{ Ultrarelativistic Heavy-Ion Collisions}}, (North-
Holland, Amsterdam, 2007).   
    
%\cite{Jaiswal:2015mxa}
\bibitem{Jaiswal:2015mxa} 
  A.~Jaiswal, B.~Friman and K.~Redlich,
  %``Relativistic second-order dissipative hydrodynamics at finite chemical potential,''
  Phys.\ Lett.\ B {\bf 751}, 548 (2015)
  %doi:10.1016/j.physletb.2015.11.018
  [arXiv:1507.02849 [nucl-th]].
  %%CITATION = doi:10.1016/j.physletb.2015.11.018;%%
  
%\cite{Haque:2014rua}
\bibitem{Haque:2014rua} 
  N.~Haque, A.~Bandyopadhyay, J.~O.~Andersen, M.~G.~Mustafa, M.~Strickland and N.~Su,
  %``Three-loop HTLpt thermodynamics at finite temperature and chemical potential,''
  JHEP {\bf 1405}, 027 (2014)
 % doi:10.1007/JHEP05(2014)027
  [arXiv:1402.6907 [hep-ph]].
  %%CITATION = doi:10.1007/JHEP05(2014)027;%%
  %126 citations counted in INSPIRE as of 13 Feb 2019
  
\bibitem{Anderson_Witting}
J.~L.~Anderson and H.~R.~Witting
%A relativistic relaxation-time for the Boltzmann equation. 
Physica \textbf{74}, 466 (1974).
    
%\cite{Jaiswal:2013npa}
\bibitem{Jaiswal:2013npa} 
  A.~Jaiswal,
  %``Relativistic dissipative hydrodynamics from kinetic theory with relaxation time approximation,''
  Phys.\ Rev.\ C {\bf 87}, 051901 (2013)
  %doi:10.1103/PhysRevC.87.051901
  [arXiv:1302.6311 [nucl-th]].
  %%CITATION = doi:10.1103/PhysRevC.87.051901;%%
  
%\cite{Jaiswal:2013vta}
\bibitem{Jaiswal:2013vta} 
  A.~Jaiswal,
  %``Relativistic third-order dissipative fluid dynamics from kinetic theory,''
  Phys.\ Rev.\ C {\bf 88}, 021903 (2013)
  %doi:10.1103/PhysRevC.88.021903
  [arXiv:1305.3480 [nucl-th]].
  %%CITATION = doi:10.1103/PhysRevC.88.021903;%%
  
 \bibitem{Landau}
L.D. Landau and E.M. Lifshitz, {\it{Fluid Mechanics}}
(Butterworth-Heinemann, Oxford, 1987).

%\cite{Bjorken:1982qr}
\bibitem{Bjorken:1982qr} 
  J.~D.~Bjorken,
  %``Highly Relativistic Nucleus-Nucleus Collisions: The Central Rapidity Region,''
  Phys.\ Rev.\ D {\bf 27}, 140 (1983).
 % doi:10.1103/PhysRevD.27.140
  %%CITATION = doi:10.1103/PhysRevD.27.140;%%
  %2936 citations counted in INSPIRE as of 13 Feb 2019

%\cite{El:2007vg}
\bibitem{El:2007vg} 
  A.~El, Z.~Xu and C.~Greiner,
  %``Thermalization of a color glass condensate and review of the 'Bottom-Up' scenario,''
  Nucl.\ Phys.\ A {\bf 806}, 287 (2008)
%  doi:10.1016/j.nuclphysa.2008.03.005
  [arXiv:0712.3734 [hep-ph]].
  %%CITATION = doi:10.1016/j.nuclphysa.2008.03.005;%%
  %56 citations counted in INSPIRE as of 13 Feb 2019
  
 %\cite{Strickland:2014pga}
\bibitem{Strickland:2014pga} 
  M.~Strickland,
  %``Anisotropic Hydrodynamics: Three lectures,''
  Acta Phys.\ Polon.\ B {\bf 45}, no. 12, 2355 (2014)
  %doi:10.5506/APhysPolB.45.2355
  [arXiv:1410.5786 [nucl-th]].
  %%CITATION = doi:10.5506/APhysPolB.45.2355;%%
  %91 citations counted in INSPIRE as of 12 Dec 2019  

%\cite{Torrieri:2008ip}
\bibitem{Torrieri:2008ip} 
  G.~Torrieri and I.~Mishustin,
  %``Instability of Boost-invariant hydrodynamics with a QCD inspired bulk viscosity,''
  Phys.\ Rev.\ C {\bf 78}, 021901 (2008)
  %doi:10.1103/PhysRevC.78.021901
  [arXiv:0805.0442 [hep-ph]].
  %%CITATION = doi:10.1103/PhysRevC.78.021901;%%
  
%\cite{Muronga:2003ta}
\bibitem{Muronga:2003ta} 
  A.~Muronga,
  %``Causal theories of dissipative relativistic fluid dynamics for nuclear collisions,''
  Phys.\ Rev.\ C {\bf 69}, 034903 (2004)
  %doi:10.1103/PhysRevC.69.034903
  [nucl-th/0309055].
  %%CITATION = doi:10.1103/PhysRevC.69.034903;%%
  %286 citations counted in INSPIRE as of 29 Mar 2020
  
%\cite{Bazow:2013ifa}
\bibitem{Bazow:2013ifa}
D.~Bazow, U.~W.~Heinz and M.~Strickland,
%``Second-order (2+1)-dimensional anisotropic hydrodynamics,''
Phys.\ Rev.\ C \textbf{90}, no.5, 054910 (2014)
%doi:10.1103/PhysRevC.90.054910
[arXiv:1311.6720 [nucl-th]].
%125 citations counted in INSPIRE as of 02 Apr 2020 
  
%\cite{BraunMunzinger:2003zd}
\bibitem{BraunMunzinger:2003zd}
P.~Braun-Munzinger, K.~Redlich and J.~Stachel,
%``Particle production in heavy ion collisions,''
%doi:10.1142/9789812795533_0008
[arXiv:nucl-th/0304013 [nucl-th]].
%557 citations counted in INSPIRE as of 02 Apr 2020

%\cite{Cleymans:2005xv}
\bibitem{Cleymans:2005xv}
J.~Cleymans, H.~Oeschler, K.~Redlich and S.~Wheaton,
%``Comparison of chemical freeze-out criteria in heavy-ion collisions,''
Phys.\ Rev.\ C \textbf{73}, 034905 (2006)
%doi:10.1103/PhysRevC.73.034905
[arXiv:hep-ph/0511094 [hep-ph]].
%526 citations counted in INSPIRE as of 02 Apr 2020

%\cite{Biswas:2020dsc}
\bibitem{Biswas:2020dsc}
D.~Biswas,
%``Centrality dependence of chemical freeze-out parameters and strangeness equilibration in RHIC and LHC,''
[arXiv:2003.10425 [hep-ph]].
%0 citations counted in INSPIRE as of 02 Apr 2020

%\cite{Bhattacharyya:2019cer}
\bibitem{Bhattacharyya:2019cer}
S.~Bhattacharyya, D.~Biswas, S.~K.~Ghosh, R.~Ray and P.~Singha,
%``Systematics of chemical freeze-out parameters in heavy-ion collision experiments,''
Phys.\ Rev.\ D \textbf{101}, no.5, 054002 (2020)
%doi:10.1103/PhysRevD.101.054002
[arXiv:1911.04828 [hep-ph]].
%1 citations counted in INSPIRE as of 02 Apr 2020
  
  
%\cite{Jaiswal:2013fc}
\bibitem{Jaiswal:2013fc} 
  A.~Jaiswal, R.~S.~Bhalerao and S.~Pal,
  %``Complete relativistic second-order dissipative hydrodynamics from the entropy principle,''
  Phys.\ Rev.\ C {\bf 87}, 021901 (2013)
  %doi:10.1103/PhysRevC.87.021901
  [arXiv:1302.0666 [nucl-th]].
  %%CITATION = doi:10.1103/PhysRevC.87.021901;%%

 \end{thebibliography}
\end{document}